\documentclass[twocolumn]{aastex62}

\usepackage{txfonts}
\usepackage{comment}
\usepackage{color,bm}
\usepackage{threeparttable}
\usepackage{hyperref}

\newcommand{\eqref}[1]{(\ref{#1})}

\shorttitle{Dusty outflows on super-puffs}
\shortauthors{Ohno \& Tanaka}

\begin{document}
%\linenumbers
%\title{Assessing the Dusty Outflows from Super-Puffs with Grain Microphysics}
%\title{Effects of Grain Growth on Radius Inflation of Super-Puffs by Dusty Outflow}
\title{Grain Growth in Escaping Atmospheres: Implications for the Radius Inflation of Super-Puffs}
%\title{Unveiling Dusty Atmospheres on Super-Puffs with Microphysical Model}

\author[0000-0003-3290-6758]{Kazumasa Ohno}
\affil{Department of Astronomy and Astrophysics, University of California Santa Cruz, 1156 High St, Santa Cruz, CA 95064, USA}
\affil{Department of Earth and Planetary Sciences, Tokyo Institute of Technology, Meguro, Tokyo, 152-8551, Japan}

\author[0000-0002-0141-5131]{Yuki A. Tanaka}
\affil{Astronomical Institute, Graduate School of Science, Tohoku University, 6-3 Aramaki, Aoba-ku, Sendai 980-8578, Japan}
\affil{Department of Earth and Planetary Sciences, Tokyo Institute of Technology, Meguro, Tokyo, 152-8551, Japan}

\begin{abstract}
Super-puffs---low-mass exoplanets with extremely low bulk density---are attractive targets for exploring their atmospheres and formation processes. Recent studies suggested that the large radii of super-puffs may be caused by atmospheric dust entrained in the escaping atmospheres. 
In this study, we investigate how the dust grows in escaping atmospheres and influences the transit radii using a microphysical model of grain growth.  
Collision growth is efficient in many cases, hindering the upward transport of dust via enhanced gravitational settling.
We find that the dust abundance in the outflow hardly exceeds the Mach number at the dust production region.
Thus, dust formed in the upper atmospheres, say at $P\la{10}^{-5} {\rm bar}$, is needed to launch a dusty outflow with high dust abundance. 
With sufficiently high dust production altitudes and rates, the dusty outflow can enhance the observable radius by a factor of $\sim2$ or even more. 
We suggest that photochemical haze is a promising candidate of high-altitude dust that can be entrained in the outflow. 
We also compute the synthetic transmission spectra of super-puff atmospheres and demonstrate that the dusty outflow produces a broad spectral slope and obscures molecular features, in agreement with featureless spectra recently reported for several super-puffs.
Lastly, using an interior structure model, we suggest that the atmospheric dust could drastically enhance the observable radius only for planets in a narrow mass range of $\sim2$--$5M_{\rm \oplus}$, in which the boil-off tends to cause total atmospheric loss. This may explain why super-puffs are uncommon despite the suggested universality of photochemical hazes.

\end{abstract}


%%%%%%%%%%%%%%%%%%%%%
\section{Introduction} \label{sec:intro}
Observational efforts in the last decade revealed the ubiquity of low-mass exoplanets with sizes between Earth and Neptune---super-Earths and sub-Neptunes---in this galaxy \citep[e.g.,][]{Mayor+11,Fressin+13,Dressing&Charbonneau15,Fulton&Petigura18}.
The Kepler mission discovered several low-mass planets whose sizes are comparable to gas giants \citep[e.g.,][]{Lissauer+11,Jontof-Hutter+14,Ofir+14,Masuda14,Mills+16,Hadden&Lithwick17,Vissapragada+20,Liang+20}.
They are called super-puffs because of their extremely low bulk density of $\rho_{\rm p}\la 0.1~{\rm g~{cm}^{-3}}$.
The large radii of super-puffs potentially imply the presence of a substantial amount of atmospheres, say $\ga10~{\rm \%}$ of the total planetary mass \citep[][]{Howe+14,Lopez&Fortney14}. 
The presence of such a massive atmosphere may offer clues to planet formation and evolution processes \citep[e.g.,][]{Ikoma&Hori12,Lee&Chiang16,Ginzburg+16}.
The low planetary gravity also facilitates atmospheric transmission spectroscopy thanks to the large atmospheric scale height, which renders super-puffs attractive targets for atmospheric characterization.
%The low planetary gravity also facilitates atmospheric transmission spectroscopy thanks to large atmospheric scale height, which renders super-puffs attractive targets for atmospheric characterization. 

While super-puffs exhibit several intriguing characteristics, they pose a challenge for the formation and evolution theories of low-mass exoplanets.
Several studies suggested that a low-mass planetary core can acquire the massive atmosphere only under the restricted conditions, namely a proper disk lifetime and cold environments \citep[e.g.,][]{Rogers+11,Ikoma&Hori12,Bodenheimer&Lissauer14,Lee&Chiang16}.
Rapid disk clearing via photoevaporation further hinders the atmospheric accretion \citep{Ogihara+20}.
Some studies suggested that super-puffs acquired their massive atmospheres at outer parts of the protoplanetary disk and then migrated to current orbits \citep{Lee&Chiang16,Chachan+21}.
Even if the massive atmosphere is successfully formed, low-mass planets are vulnerable to intense atmospheric escape after disk disappation \citep[e.g.,][]{Ikoma&Hori12,Stokl+15,Chen&Rogers16,Owen&Wu16,Ginzburg+16,Fossati+17,Wang&Dai18,Kubyshkina+18}.
In fact, \citet{Cubilios+17} reported that about $15\%$ of low-mass planets ($<30M_{\rm \oplus}$) exhibits too high a mass-loss rate to sustain their atmospheres.

Another enigma is raised by the featureless transmission spectra of atmospheres recently reported for several super-puffs.
For example, WASP-107b, with a mass similar to Neptune ($30M_{\rm \oplus}$) and a radius similar to Jupiter ($10~{\rm R_{\rm \oplus}}$) \citep{Anderson+17,Piaulet+20}\footnote{WASP-107b may be better classified as a sub-Saturn rather than a super-puff. In this paper, we use the nomenclature of "super-puff" as a loose meaning of planets with masses approximately lower than that of Neptune and bulk densities lower than that of Saturn ($0.69~{\rm g~{cm}^{-3}}$).}, exhibits a muted H$_2$O feature that is hardly explained by a cloud-free atmosphere \citep{Kreidberg+18}.
Kepler-51b, with a lower mass of $3.69M_{\rm \oplus}$ and a radius of $6.89~{\rm R_{\rm \oplus}}$ \citep[updated from][]{Masuda14}, also shows the featureless spectrum that is best fitted by a flat line \citep{Libby-Roberts+20}.
The same result was also reported for Kepler-51d with a mass of $5.70M_{\rm \oplus}$ and a radius of $9.46~{\rm R_{\rm \oplus}}$.
Most recently, \citet{Chachan+20} reported the featureless spectrum for Kepler-79d, with a mass of $5.30M_{\rm \oplus}$ and a radius of $7.15~{\rm R_{\rm \oplus}}$ \citep[updated from][]{Jontof-Hutter+14}.
These results are surprising because it was expected that large atmospheric scale height of super-puffs sculpt prominent atmospheric features in their spectra.

Recent studies suggested that the observed radii of super-puffs may be much larger than their actual radii due to some reasons.
Several studies attributed the large radii to the presence of atmospheric dust \footnote{In this study, we use the term of ``dust'' to express any solid and liquid particles floating in atmospheres following the nomenclature of \citet{Wang&Dai19}. Thus,  ``dust'' can be replaced by ``aerosols'' in this paper. We note that some literature only used the ``dust'' to express solid particles lifted from the ground, such as the dust devil of Mars \citep[see e.g.,][]{Gao+21}. 
} that elevates the pressure level probed by the transit observation to the upper atmospheres.
The idea was originally invoked by \citet{Lammer+16} to remedy the mass-loss timescale of a low-density hot Neptune CoRoT-24b.
In particular, \citet{Wang&Dai19} suggested that atmospheric escape from super-puffs can blow up dust to upper atmospheres, leading to the enhancement of the observable radius over the actual radius by a factor of $\sim3$. 
They succeeded in explaining the large radius of Kepler-51b with a reasonable atmospheric mass-loss timescale when the dust-to-gas mass ratio is as high as $10^{-2}$.
Similarly, \citet{Gao&Zhang20} suggested that photochemical haze formed at high altitudes can enhance the observable radius as well when the planet is young ($0.1$--$1~{\rm Gyr}$), low-mass ($<4M_{\oplus}$), and warm ($T_{\rm eq}>500~{\rm K}$) \citep[see also][]{Kawashima+19}.
Alternatively, several studies suggested that the low bulk density may imply the presence of a circumplanetary ring \citep{Piro&Vissapragada19,Akinsanmi+20}. 
\citet{Millholland19} suggested that external interior heating, such as tidal heating, can explain the large radii of super-puffs, though additional mechanisms must be invoked to explain the featureless transmission spectra.

In this study, we investigate the dusty outflow scenario originally proposed by \citet{Wang&Dai19} in detail. 
\citet{Wang&Dai19} assumed a constant dust-to-gas mass ratio ($\sim{10}^{-2}$) and particle sizes ($0.001~{\rm {\mu}m}$) in escaping atmospheres.
We study how the dust size and abundance evolve in escaping atmospheres using a microphysical model of grain growth.
Our study is similar to a recent study of \citet{Gao&Zhang20} that also used a microphysical model to simulate the photochemical haze formation on super-puffs.
Instead of calculating the haze production rate from first principles like them, we parameterize dust formation processes by varying the dust production rate and altitude. 
This approach enables us to discuss what kinds of atmospheric dust are responsible to the radius enhancement without assuming photochemical hazes alone.
Thus, our study complements that of \citet{Gao&Zhang20}.

We will also attempt to give an explanation for why super-puffs are uncommon.
Anomalously low-density planets are only $\sim15\%$ of the whole sample of low-mass planets \citep{Cubilios+17}.
One possible explanation for the rarity of super-puffs is that they originate in special formation conditions \citep{Lee&Chiang16}.
However, this topic has been less discussed in the scenario that explains super-puffs by enhanced transit radius.
\citet{Wang&Dai19} stated that the dusty outflow tends to work for low-mass planets with $\la10M_{\rm \oplus}$ owing to enhanced atmospheric escape, but this unlikely explains that the majority of low-mass planets are not super-puffs.
We will show that the radius enhancement can occur only in a limited planetary mass range, in which the planet is always on the verge of total atmospheric loss.

The organization of this paper is as follows.
In Section \ref{sec:method}, we introduce the methodology adopted in this study.
In Section \ref{sec:results}, we show how dust abundance and size distributions vary with dust production rate, production altitude, and mass-loss timescales in escaping atmospheres. 
We also provide a simple analytical theory that predicts the dust abundance in the outflow.
In Section \ref{sec:observation}, we investigate how the dusty outflow affects the observable planetary radius and atmospheric transmission spectra.
Then, we discuss what properties of atmospheric dust are needed to enhance the observable transit radius.
In Section \ref{sec:MR}, we discuss  why super-puffs are a rare population of low-mass planets. 
In Section \ref{sec:discussion}, we discuss implications for young exoplanet observations, impacts of particle porosity evolution on the results of present study, and the observational diagnosis of circumplanetary rings.
In Section \ref{sec:summary}, we summarize our findings.

%%%%%%%%%%%%%%%%%%%%%%%%%%%%%%%%%
\section{Method} \label{sec:method}
\subsection{Overview}
We investigate how dust particles evolve in escaping atmospheres using a microphysical model of \citet{Ohno+20b} that was originally used to study haze formation on Triton.
The model adopts a 1D Eulerian framework and simulates the evolution of a particle size distribution at each altitude by taking into account vertical transport and growth via collision and condensation.
Whether or not the condensation growth takes place depends on an atmospheric thermal structure, which differs from planet to planet.
To keep generality, we switch off the condensation growth in this study.
Although the model was designed to simulate the porosity evolution of dust particles, we assume compact spherical particles for the sake of simplicity. We will show a few simulation results with particle porosity evolution in Section \ref{sec:porosity}.

Our interest lies in whether particle sizes remain small enough such that the atmospheric escape can blow up dust.
One can estimate a threshold size above which the outflow cannot transport the dust as follows.
Assuming dust much smaller than the gas mean free path, the terminal settling velocity of dust is approximated by \citep[e.g.,][]{Woitke&Helling03}
\begin{equation}\label{eq:v_eps}
     %v_{\rm d}\sim \frac{\rho_{\rm p}GM_{\rm p}}{\rho_{\rm g}c_{\rm s}r^2}a, 
     v_{\rm d}\approx \frac{\rho_{\rm d}GM_{\rm p}}{\rho_{\rm g}r^2\sqrt{8k_{\rm B}T/\pi m_{\rm g}}}a,  
\end{equation}
where $\rho_{\rm d}$ is the dust internal density, $M_{\rm p}$ is the planetary mass, $\rho_{\rm g}$ is the gas density, $r$ is the radial distance from the planet center, $T$ is the atmospheric temperature, $m_{\rm g}$ is the mean mass of gas particles, and $a$ is the radius of a dust particle.
On the other hand, the outflow velocity of the atmosphere $v_{\rm g}$ can be estimated from the mass conservation of
\begin{equation}\label{eq:flux_conserve}
    %4\pi r^2 \rho_{\rm g}v_{\rm g}=\dot{M},
    v_{\rm g}=\frac{\dot{M}}{4\pi r^2 \rho_{\rm g}},
\end{equation}
where $\dot{M}$ is the atmospheric mass-loss rate.
The atmospheric escape cannot blow up the dust when the terminal velocity is faster than the outflow velocity.
Equating Equations \eqref{eq:v_eps} and \eqref{eq:flux_conserve}, we obtain the threshold size below which the atmospheric escape can blow up dust:
\begin{eqnarray}\label{eq:a_cri}
    a_{\rm cri}&\approx&\frac{\dot{M}\sqrt{8k_{\rm B}T/\pi m_{\rm g}}}{4\pi \rho_{\rm d}GM_{\rm p}}\\
    \nonumber
    &\sim&0.1~{\rm \mu m}\left(\frac{\tau_{\rm loss}}{1~{\rm Gyr}}\right)^{-1}\left(\frac{T}{500~{\rm K}}\right)^{1/2}\left(\frac{m_{\rm g}}{2.35~{\rm amu}}\right)^{-1/2} \left(\frac{\rho_{\rm d}}{1~{\rm g~{cm}^{-3}}}\right)^{-1},
\end{eqnarray}
where we have defined the mass-loss timescale $\tau_{\rm loss}\equiv M_{\rm p}/\dot{M}$.
Note that the threshold size is independent of a radial distance \citep[see also][]{Wang&Dai18,Wang&Dai19}.
Atmospheric escape can blow up larger particles for shorter $\tau_{\rm loss}$, whereas the mass-loss timescale should be at least comparable to the system age to sustain the atmosphere.
For the typical system ages of Kepler planets \citep[${\sim}3~{\rm Gyr}$,][]{Berger+20}, the dust should be smaller than $\sim0.1~{\rm \mu m}$.
In reality, the atmospheric lifetime would be shorter than $\tau_{\rm loss}$ defined above, as the atmospheric mass fraction of low-mass planets is typically $0.01$--$0.3$ \citep{Lopez&Fortney14}.

%%%%%%%%%%%%%%%%%%%%%%%%%%%%%%%%%%%%%%%%%%%%%
\subsection{Microphysics of Grain Growth}\label{sec:method_microphysics}
The evolution of the particle size distribution is described by the Smoluchowski equation, given by
\begin{eqnarray}\label{eq:basic}
\nonumber
\frac{\partial n(m)}{\partial t}&=&\frac{1}{2}\int_{\rm 0}^{\rm m}~K(m',m-m')~n(m')~n(m-m')~dm' \\
\nonumber
&&-n(m)\int_{\rm 0}^{\rm \infty}K(m,m')~n(m')~dm'\\
\nonumber
&&+\frac{1}{r^2}\frac{\partial}{\partial r}\left[ r^2\left(\rho_{\rm g}K_{\rm z}\frac{\partial}{\partial r}\left( \frac{n(m)}{\rho_{\rm g}}\right)-(v_{\rm g}-v_{\rm d})n(m)\right)\right]\\
&&-\frac{1}{4\pi r^2m}\frac{\partial}{\partial m}\left(\frac{\partial \dot{M}_{\rm dust}}{\partial r}\right)
\end{eqnarray}
where $n(m)dm$ is the number density of particles with masses between $m$ and $m+dm$, $K(m_{\rm 1},m_{\rm 2})$ is the collision kernel describing the collision rate between particles with mass of $m_{\rm 1}$ and $m_{\rm 2}$, $K_{\rm z}$ is the eddy diffusion coefficient, and $\dot{M}_{\rm dust}$ is the total dust production rate in the atmosphere.
{  In the fiducial simulations, we omit the eddy diffusion to focus on the ability of atmospheric escape to transport dust upward.
We will examine the effects of the eddy diffusion transport in Section \ref{sec:Kzz}.}
%We assume $K_{\rm z}={10}^{3}~{\rm {m}^2~s^{-1}}$, which is comparable to the value estimated for GJ1214b \citep{Charnay+15a} whose equilibrium temperature is close to super-puff Kepler-51b.

{  
We have assumed perfect sticking for dust collisions in Equation \eqref{eq:basic}, which has been a common assumption in previous microphysical models of atmospheric dust.
In reality, there are a number of possible outcomes for dust collisions, such as fragmentation, bouncing, and erosion \citep[e.g.,][]{Guttler+10}.
For microphysical models including collision fragmentation, we refer readers to theoretical studies of grain growth in protoplanetary disks \citep[e.g.,][]{Brauer+08,Birnstiel+10}.
}

The terminal velocity depends on a gas drag law that varies with particle size, ambient gas density, and the velocity itself.
We use the formula of \citet{Ohno&Okuzumi17} that is applicable to multiple gas drag regimes, given by
\begin{equation}\label{eq:vt}
v_{\rm d} =\frac{2ga^2\rho_{\rm d}}{9\eta}\mathcal{B}(a) \left[ 1+\left( \frac{0.45g a^3\rho_{\rm g}\rho_{\rm p}}{54\eta^2}\right)^{2/5}\right]^{-5/4},
\end{equation}
where $g$ is the gravitational acceleration, $\eta$ is the dynamic viscosity, and $\mathcal{B}$ is the Cunningham slip correction factor, given by \citep[e.g.,][]{Seinfeld&Pandis98}
\begin{equation}
    \mathcal{B}=1+\frac{l_{\rm g}}{a}\left[1.257+0.4\exp{\left( -1.1\frac{a}{l_{\rm g}}\right)}\right],
\end{equation}
where $l_{\rm g}$ is the gas mean free path.
The slip correction factor accounts for the transition from laminar flow (Stokes) to free molecular flow (Epstein) regimes, and the bracket part of Equation \eqref{eq:vt} accounts for the transition from the laminar flow to turbulent flow (Newton) regimes \citep[for details, see the Appendix of][]{Ohno&Okuzumi17}.

The collision growth is driven by the thermal Brownian motion and gravitational settling.
We calculate the kernel as a root-sum-square of each component
\begin{equation}
    K(m_{\rm 1},m_{\rm 2})=\sqrt{K_{\rm B}^2+K_{\rm G}^2},
\end{equation}
where the $K_{\rm B}$ and $K_{\rm G}$ are the kernels for collisions driven by the thermal Brownian motion and gravitational settling, respectively. 
The former is given by \citep[Eq. (15.33) of][]{Jacobson05},
\begin{eqnarray}\label{eq:K_coag}
    K_{\rm B}(m_{\rm 1},m_{\rm 2})&=&4\pi(a_{\rm 1}+a_{\rm 2})(D_{\rm p,1}+D_{\rm p,2})\\
    \nonumber
    &&\times
    \left( \frac{a_{\rm 1}+a_{\rm 2}}{a_{\rm 1}+a_{\rm 2}+\sqrt{\delta_{\rm 1}^2+\delta_{\rm 2}^2}} + \frac{4(D_{\rm p,1}+D_{\rm p,2})}{\sqrt{\overline{v}_{\rm 1}^2+\overline{v}_{\rm 2}^2}(a_{\rm 1}+a_{\rm 2})} \right)^{-1}.
\end{eqnarray}
Here, $\overline{v}_{\rm i}=\sqrt{8k_{\rm B}T/\pi m_{\rm i}}$ is the mean thermal velocity of particles, and $D_{\rm p,i}$ is the particle diffusion coefficient given by
\begin{equation}
    D_{\rm p,i}=\frac{k_{\rm B}T}{6\pi a_{\rm i}\eta}\mathcal{B}.
\end{equation}
The term $\delta_{\rm i}$ is the mean distance from the center of a sphere reached by particles leaving the particle surface, given by
\begin{equation}
\delta_{\rm i}=\frac{(2a_{\rm i}+\lambda_{\rm i})^3-(4a_{\rm i}^2+\lambda_{\rm i}^{2})^{3/2}}{6a_{\rm i}\lambda_{\rm i}}-2a_{\rm i},
\end{equation}
where $\lambda_{\rm i}=8D_{\rm p,i}/\pi\overline{v}_{\rm i}$ is the particle mean free path.
The collision kernel for the gravitational settling is given by
\begin{equation}
    K_{\rm G}(m_{\rm 1},m_{\rm 2})=E_{\rm coll}\pi(a_{\rm 1}+a_{\rm 2})^2|v_{\rm d}(m_{\rm 1})-v_{\rm d}(m_{\rm 2})|,
\end{equation}
where $E_{\rm coll}$ is the collision efficiency accounting for the fact that the particles cannot attach to each other if the motion of the smaller particle is tightly coupled to gas stream lines.
We use a smoother analytic function of \citet{Guillot14}, given by
\begin{equation}
    E_{\rm coll}=\max{[0,1-0.42{\rm St}^{-0.75}]}.
\end{equation}
The efficiency is characterized by the Stokes number St, the ratio of particle stopping time to crossing time, defined as
\begin{equation}
    {\rm St}\equiv \frac{v_{\rm d}(a_{\rm 2})}{g}\frac{v_{\rm d}(a_{\rm 1})-v_{\rm d}(a_{\rm 2})}{a_{\rm 1}},
\end{equation}
where $a_{\rm 1}>a_{\rm 2}$.
The smaller particle is tightly coupled to the stream line and hardly collides with the larger particle when ${\rm St}\ll1$.
This effect should work only when the ambient gases behave as a continuum \citep{Rossow78}. 
Thus, we assume $E_{\rm coll}=1$ if the gas mean free path is larger than $a_{\rm 1}$.

We parameterize the dust production profile using a log-normal distribution given by 
\begin{equation}
    \frac{\partial}{\partial m}\left(\frac{\partial \dot{M}_{\rm dust}}{\partial r}\right)=\frac{\partial P}{\partial r}\frac{f_{\rm dust}\dot{M}\delta(m-m_{\rm 0})}{\sigma P\sqrt{2\pi}}\exp{\left[ -\frac{\ln^{2}{(P/P_{\rm 0})}}{2\sigma^2}\right]},
    %S(m,P)=-\frac{dP}{dr}\frac{f_{\rm dust}\dot{M}/4\pi r^2 m_{\rm 0}}{\sigma P\sqrt{2\pi}}\delta(m-m_{\rm 0})\exp{\left( -\frac{(\log{(P/P_{\rm 0})})^2}{2\sigma^2}\right)},
\end{equation}
where $f_{\rm dust}\equiv \dot{M}_{\rm dust}/\dot{M}$ is the ratio of the total dust production rate to the atmospheric mass-loss rate, $P_{\rm 0}$ is the atmospheric pressure of dust production, $\sigma$ is the width of the distribution set to $\sigma=0.5$, and $m_{\rm 0}$ is the mass of particles in the smallest mass grid.
The dust mass mixing ratio approaches to $f_{\rm dust}$ at $P\ll P_{\rm 0}$ when the particle settling is negligible.
The dust production pressure is related to the type of atmospheric dust; for instance, low $P_{\rm 0}$ mimics high-altitude dust formation such as photochemical haze, while high $P_{\rm 0}$ mimics low-altitude dust formation such as condensation clouds.
Several studies adopted a column-integrated production rate instead of $f_{\rm dust}$ \citep[e.g.,][]{Adams+19,Ohno&Kawashima20}.
The column-integrated rate $F_{\rm top}$ can be expressed by
\begin{eqnarray}
    F_{\rm top}&\approx&\frac{f_{\rm dust}\dot{M}}{4\pi r(P_{\rm 0})^2},\\
    \nonumber
    &\approx& 3.8\times{10}^{-12}~{\rm g~{cm}^{-2}~s^{-1}}\left( \frac{f_{\rm dust}}{{10}^{-4}}\right)\left( \frac{\tau_{\rm loss}}{1~{\rm Gyr}}\right)^{-1}\left( \frac{g}{{10}~{\rm m~s^{-2}}}\right),
\end{eqnarray}
where $g$ is the surface gravity measured at the dust production region.
In the context of photochemical haze formation, photochemical models predict the column-integrated rate of $\sim{10}^{-11}$--${10}^{-13}~{\rm g~{cm}^{-2}~s^{-1}}$ for a GJ1214b-like planet \citep{Kawashima&Ikoma19,Lavvas+19}.
When the production rate is limited by the transport flux of precursors as assumed in \citet{Adams+19} and \citet{Gao+20}, the $f_{\rm dust}$ would be comparable to the mass mixing ratio of precursor molecules ($\sim3\times{10}^{-3}$ in the case of CH$_4$ in solar composition atmospheres at $<1000~{\rm K}$ {according to Figure 3 of \citet{Woitke+18}}) multiplied by a reduction factor accounting for conversion efficiency. 

%%%%%%%%%%%%%%%%%%%%%%%%%%%%%%%%%%%%
\subsection{Atmospheric Escape Model}
We calculate the outflow velocity $v_{\rm g}$ using an isothermal Parker wind model \citep{Parker58}.
The Parker wind mass-loss has also been called ``boil-off'' in the exoplanet community \citep{Owen&Wu16,Fossati+17}, and the outflow is driven solely by the pressure gradient from the deep atmosphere.
Though several other heating sources can drive the outflow, such as photoionization of hydrogen caused by stellar X-ray and extreme ultraviolet irradiation \citep[the so-called XUV-driven escape; e.g.,][]{Lammer+03,Yelle+04,Tian+05,Murray-Clay+09}, photoelectric heating of dust by stellar far-ultraviolet irradiation \citep{Mitani+20}, and the dissipation of magnetohydronydamic waves \citep{Tanaka+14,Tanaka+15}, it has been suggested that the boil-off dominates over others for low-mass and/or inflated planets, such as super-puffs \citep{Fossati+17,Kubyshkina+18}.

From the mass continuity and momentum equations with a constant sound speed, the Parker wind model gives an analytic equation that solely depends on the outflow velocity  \citep[e.g.,][]{Parker58,Wang&Dai19}
\begin{equation}\label{eq:parker}
    \left( \frac{v_{\rm g}}{c_{\rm s}}\right)\exp{\left(-\frac{v_{\rm g}^2}{2c_{\rm s}^2}\right)}=\left( \frac{r}{r_{\rm s}}\right)^{-2}\exp{\left(\frac{3}{2} - \frac{2r_{\rm s}}{r} \right)},
\end{equation}
where $c_{\rm s}=\sqrt{k_{\rm B}T/m_{\rm g}}$ is the isothermal sound speed, and $r_{\rm s}$ is the sonic radius given by
\begin{equation}
    r_{\rm s}=\frac{GM_{\rm p}}{2c_{\rm s}^2}.
\end{equation}
Equation \eqref{eq:parker} can be numerically solved to obtain $v_{\rm g}$.
The outflow velocity is much slower than the sound speed at subsonic regions, and thus the velocity at $r<r_{\rm s}$ can be approximated by
\begin{equation}\label{eq:vg_appro}
    v_{\rm g}\approx c_{\rm s}\left( \frac{r}{r_{\rm s}}\right)^{-2}\exp{\left(\frac{3}{2} - \frac{2r_{\rm s}}{r} \right)}.
\end{equation}
This equation yields the mass-loss rate $\dot{M}$ for the Parker wind, given by
\begin{equation}\label{eq:M_dot}
    \dot{M}\approx 4\pi r_{\rm s}^2c_{\rm s}\rho_{\rm g}(r_{\rm s}).%\exp{\left(\frac{3}{2} - \frac{2r_{\rm s}}{r} \right)}.
\end{equation}
Strictly speaking, the isothermal condition does not hold in general: as the radial distance increases, the temperature decreases via adiabatic cooling \citep[][]{Fossati+17,Kubyshkina+18} or increases due to heating of dust entrained in the outflow \citep{Wang&Dai18,Wang&Dai19}.
Nonetheless, \citet{Wang&Dai19} obtained similar results for the isothermal model and sophisticated hydrodynamic simulations.
Motivated by their results, we decided to adopt a relatively simple isothermal model in this study.

We note that the mass-loss rate given by Equation \eqref{eq:M_dot} should be regarded as an upper limit.
In reality, the mass-loss rate cannot exceed the maximum rate determined by the total energy of the bolometric stellar flux.
Such a bolometric flux limited mass-loss rate is given by \citep{Wang&Dai18}
\begin{equation}
    \dot{M}_{\rm max}\sim \epsilon\left( \frac{L_{\rm *}}{4\pi a_{\rm orb}^2}\right)\pi R_{\rm p}^2\left( \frac{c_{\rm s}^2}{2}\right)^{-1},
    %\dot{M}= 4\pi R_{\rm p}^2\sigma_{\rm SB}(T_{\rm eq}^4-T^4)\left(\frac{c_{\rm s}^2}{2}\right)^{-1},
\end{equation}
where $L_{\rm *}$ is the stellar luminosity, $a_{\rm orb}$ is the orbital distance, and $R_{\rm p}$ is the planetary radius measured at the radiative-convective boundary. 
We also introduce an energy conversion efficiency of $\epsilon$, as the planet likely radiates away some fraction of received energy.
Note that energy conservation implies the atmospheric temperature of $T=(1-\epsilon)^{1/4}T_{\rm eq}$, where $T_{\rm eq}=(L_{\rm *}/16\pi \sigma_{\rm SB}a_{\rm orb}^2)^{1/4}$ is the equilibrium temperature for zero bond albedo and $\sigma_{\rm SB}$ is the Stefan-Boltzmann constant.
The timescale of bolometric energy limited mass-loss is then given by
\begin{eqnarray}\label{eq:bolo}
    \tau_{\rm loss,min}&\sim& \frac{M_{\rm p}c_{\rm s}^2}{8\pi R_{\rm p}^2\sigma_{\rm SB}T_{\rm eq}^4\epsilon}\\
    \nonumber
    &\sim& 1~{\rm Myr}~\frac{(1-\epsilon)^{1/4}}{\epsilon}\left( \frac{g}{10~{\rm m~s^{-2}}}\right)\left( \frac{T_{\rm eq}}{500~{\rm K}}\right)^{-3}\left( \frac{m_{\rm g}}{2.35~{\rm amu}}\right)^{-1}.
\end{eqnarray}
We will perform simulations only for $\tau_{\rm loss}\ge100~{\rm Myr}$. 
Thus, the bolometric flux is always high enough to drive the assumed isothermal wind, unless the conversion efficiency $\epsilon$ is extremely low.

%%%%%%%%%%%%%%%%%%%%%%%%%%%%%%%%%%%
\begin{figure*}[t]
\centering
\includegraphics[clip, width=\hsize]{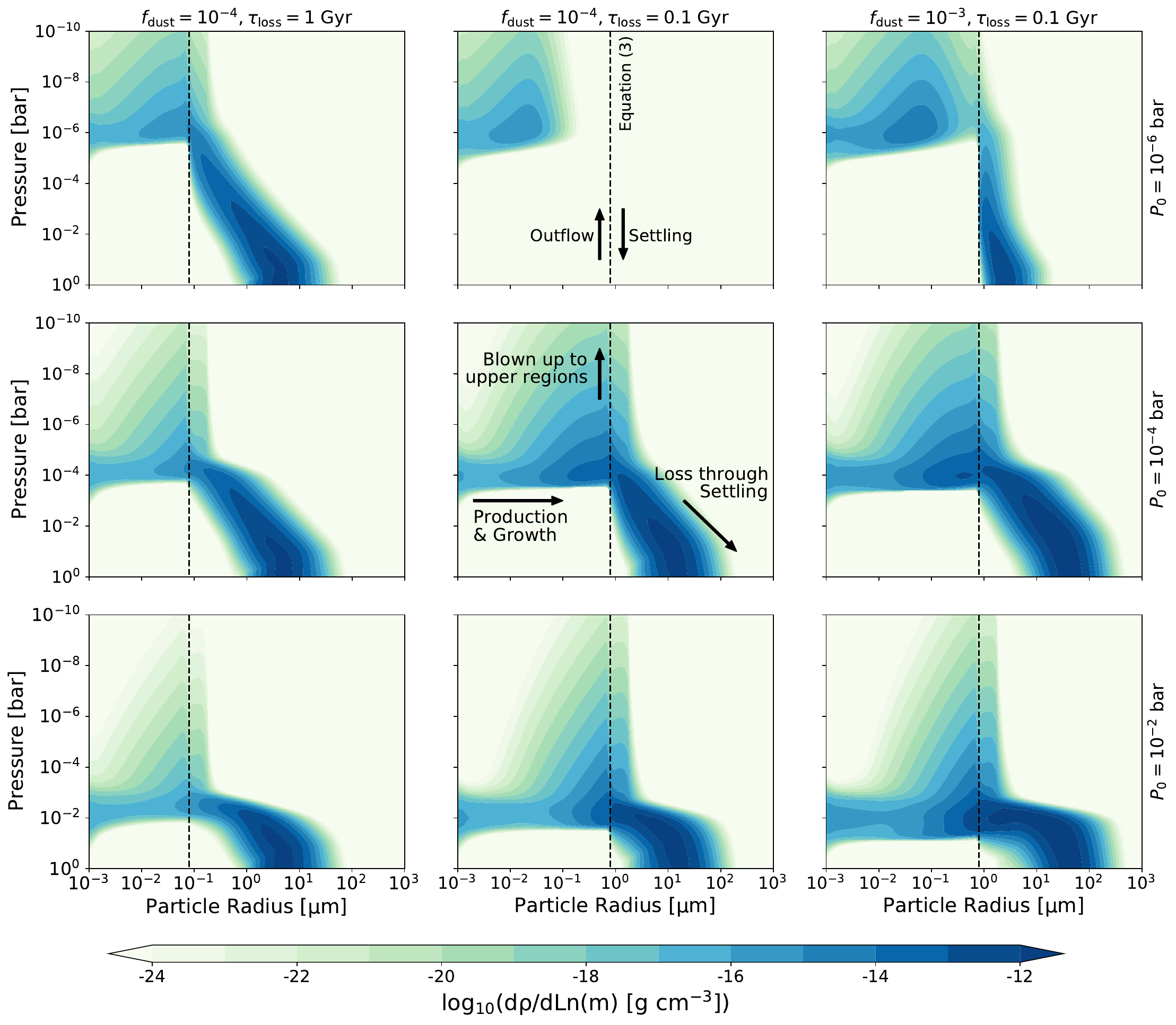}
\caption{Vertical mass distributions of aerosol particles $m^2n(m)$ (colorscale) in escaping atmospheres. The vertical and horizontal axes are atmospheric pressure and particle radius. From top to bottom, each row exhibits the distribution for $P_{\rm 0}={10}^{-6}$, ${10}^{-4}$, and ${10}^{-2}~{\rm bar}$, respectively. The left and middle columns show the distributions for $f_{\rm dust}={10}^{-4}$ with the mass-loss timescale of $\tau_{\rm loss}=1$ and $0.1~{\rm Gyr}$, while the right column shows the distributions for $f_{\rm dust}={10}^{-3}$ and $\tau_{\rm loss}=0.1~{\rm Gyr}$. In each panel, the black dashed lines show the threshold radius at which the particle terminal velocity equals the outflow velocity, given by Equation \eqref{eq:a_cri}.
}
\label{fig:dust_distribution}
\end{figure*}

\begin{figure*}[t]
\centering
\includegraphics[clip, width=\hsize]{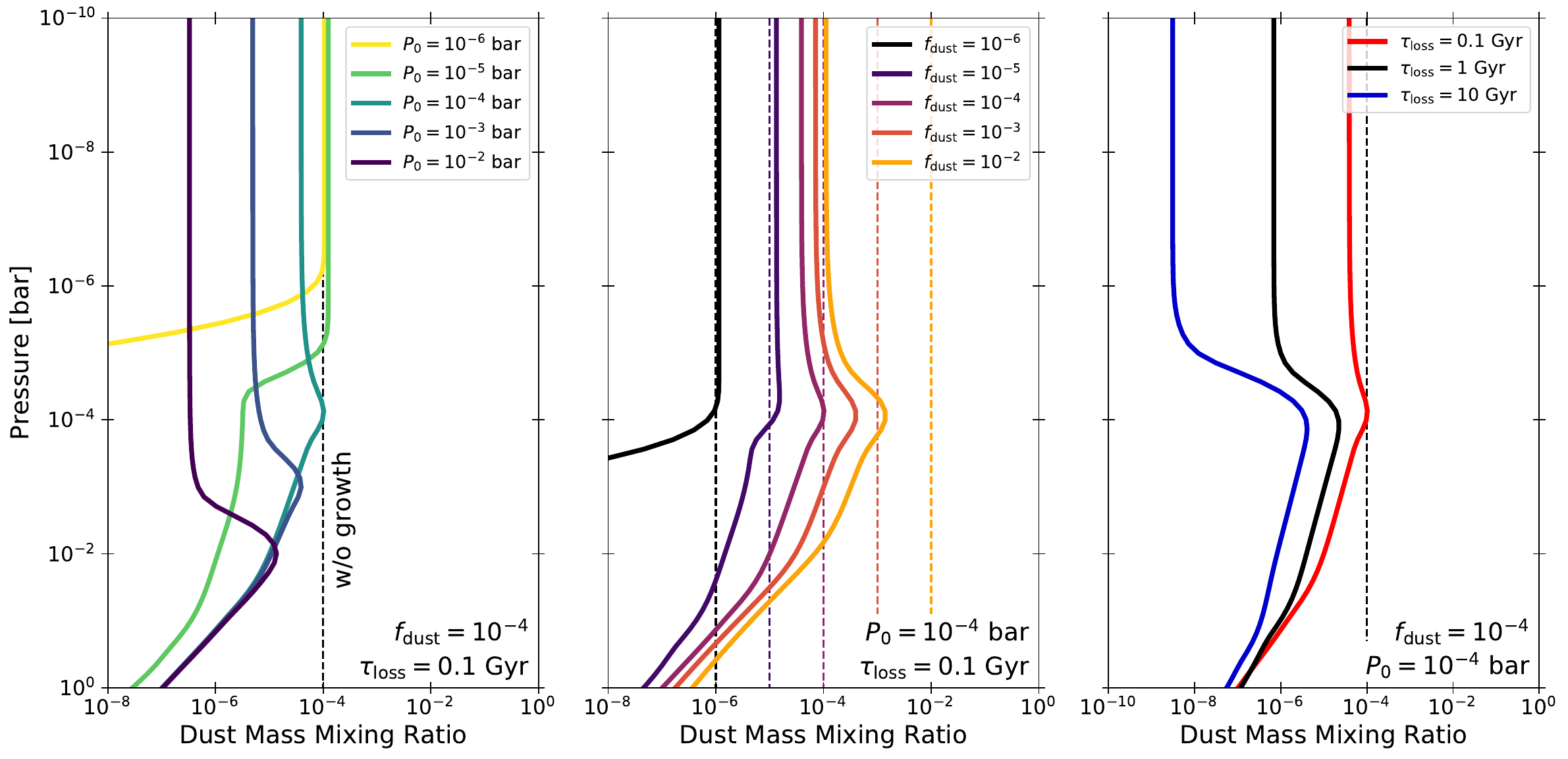}
\caption{Vertical mass distributions of aerosol particles. The vertical and horizontal axes are atmospheric pressure and particle mass mixing ratio.
Different colored lines show the distributions for different aerosol production altitude $P_{\rm 0}$ in the left panel, different production rate $f_{\rm dust}$ in the middle panel, and different mass-loss timescale $\tau_{\rm loss}$ in the right panel. We set $P_{\rm 0}=10^{-4}~{\rm bar}$ in the middle and right panels, $f_{\rm dust}=10^{-4}$ in the left and right panels, and $\tau_{\rm loss}=0.1~{\rm Gyr}$ in the left and middle panels.%Different colored lines show the distributions for different aerosol production altitudes $P_{\rm 0}$. The left and right panels show the distributions for the production rate of $f_{\rm dust}={10}^{-4}$ with $\tau_{\rm loss}=10$ and $0.1~{\rm Gyr}$, while the right panel shows the distribution for $f_{\rm dust}={10}^{-2}$ with $\tau_{\rm loss}=0.1~{\rm Gyr}$. 
The black dashed lines denote $f_{\rm dust}$, the mass mixing ratio expected for dusty outflow without particle settling.}
\label{fig:density_distribution}
\end{figure*}

%%%%%%%%%%%%%%%%%%%%%%%%%%%%%%%%%%%
%%%%%%%%%%%%%%%%%%%%%%%%%%%%%%%%%%%%
\subsection{Numerical Procedures}
We solve Equation \eqref{eq:basic} until the system reaches a steady state.
The radial coordinate is divided by spatial grids of $dr=0.3r^2/r_{\rm s}$.
The mass coordinate is divided into linearly spaced bins $m_{\rm k}=km_{\rm 0}$ for $m_{\rm k}<N_{\rm bd}m_{\rm 0}/2$ and logarithmically spaced bins $m_{\rm k}=m_{\rm k-1}{10}^{1/N_{\rm bd}}$ for $m_{\rm k}\geq N_{\rm bd}m_{\rm 0}/2$, where we adopt the mass resolution of $N_{\rm bd}=5$, i.e., $m_{\rm k}/m_{\rm k-1}{\approx} 1.58$.
We set the smallest mass grid $m_{\rm 0}$ assuming the radius of $a_{\rm 0}=0.001~{\rm {\mu}m}$ and particle internal density of $\rho_{\rm d}=1~{\rm g~{cm}^{-3}}$, as similar to \citet{Wang&Dai19}.
For mass grids with $v_{\rm g}>v_{\rm d}$ ($v_{\rm g}<v_{\rm d}$), we set zero incoming flux conditions at the lower (upper) boundary and allow particles to flow out at another boundary of the computation domain freely. 
{ 
In simulations including eddy diffusion (Section \ref{sec:Kzz}), which represents vertical mixing by atmospheric circulation, we adopt zero diffusion flux at the upper boundary, as the circulation does not escape from the planet.
For the lower boundary, we compute the downward diffusion flux by assuming a zero particle number density at the bottom of the computational domain, since particles are eventually thermally decomposed at deep hot atmospheres.
}

We suppose a hypothetical planet with a mass of $3.5M_{\rm Earth}$ and an atmospheric temperature of $500~{\rm K}$, similar to Kepler-51b.
A solar composition atmosphere with a mean molecular mass of $m_{\rm g}=2.35~{\rm amu}$ is assumed for all simulations.
We vary the mass-loss timescale $\tau_{\rm loss}{\equiv}M_{\rm p}/\dot{M}$ as a free parameter.
For given $\tau_{\rm loss}$, we calculate the atmospheric density structure from the outflow velocity (Equation \ref{eq:parker}) and mass conservation law (Equation \ref{eq:flux_conserve}).
We also vary the dust production rate $f_{\rm dust}$ and pressure $P_{\rm 0}$ as free parameters.

%%%%%%%%%%%%%%%%%%%%%%%%%%%%%%%%%
\section{Results}\label{sec:results}

\subsection{Dust Size Distributions in Escaping Atmospheres}\label{sec:result1}
There are two distinct fates of the dust formed in atmospheres: upward transport by atmospheric outflow and downward transport by gravitational settling. 
Figure \ref{fig:dust_distribution} shows the vertical mass distributions $m^2n(m)$ for different dust production pressure $P_{\rm 0}$, production rate $f_{\rm dust}$, and mass-loss timescale $\tau_{\rm loss}$.
The upward transport is always dominant for particles smaller than the threshold size given by Equation \ref{eq:a_cri}, while the downward transport becomes dominant when the particle sizes exceed the threshold size.
The settling particles gradually increase in size as they sink to a deeper atmosphere, which is similar to photochemical haze formation \citep[e.g.,][]{Lavvas&Koskinen17,Kawashima&Ikoma18,Adams+19,Gao&Zhang20,Ohno&Kawashima20}.
On the other hand, the outflowing dust keeps nearly constant sizes that are mostly determined by collision growth around the dust production region.
While \citet{Wang&Dai19} fixed the particle sizes to $0.001~{\rm {\mu}m}$, we find that the dust usually grows to much larger sizes via collision growth in escaping atmospheres.

It is intuitively understandable that the more intense atmospheric escape is, the more dust is blown up to upper atmospheres.
For example, in the case of $P_{\rm 0}={10}^{-6}~{\rm bar}$ and $f_{\rm dust}={10}^{-4}$, the majority of particles settle down to the planet for $\tau_{\rm loss}=1$ Gyr, while all particles are transported upward for $\tau_{\rm loss}=0.1$ Gyr (left and middle top panels of Figure \ref{fig:dust_distribution}).
This is simply because the more intense outflow can transport larger particles (Equation \ref{eq:a_cri}).
Moreover, the intense outflow can blow up dust particles before they grow into large sizes, further enhancing the upward transport.

The high dust production rate $f_{\rm dust}$ leads to efficient particle growth, which induces subsequent gravitational settling.
In the case of $P_{\rm 0}={10}^{-6}~{\rm bar}$ and $\tau_{\rm loss}=0.1~{\rm Gyr}$, all dust particles are transported to upper atmospheres for $f_{\rm dust}={10}^{-4}$, whereas a substantial amount of dust settle down to the planet for $f_{\rm dust}={10}^{-2}$ (middle and right top panels of Figure \ref{fig:dust_distribution}).
This demonstrates the importance of particle growth, especially for the outflow with high dust abundance.

The dust production height is another critical parameter that controls how much the outflow can blow up dust to upper atmospheres.
For $f_{\rm dust}={10}^{-4}$ and $\tau_{\rm loss}=0.1~{\rm Gyr}$, all dust particles are transported upward for $P_{\rm 0}={10}^{-6}~{\rm bar}$, whereas substantial amounts of dust settle down to the planet for $P_{\rm 0}={10}^{-5}$ and ${10}^{-4}~{\rm bar}$  (middle column of Figure \ref{fig:dust_distribution}).
Thus, the outflow can blow up more dust when the dust is formed at higher altitudes.
This is because the outflow velocity rapidly increases with increasing radial distance.
Moreover, particle growth tends to be more efficient in deeper atmospheres where the dust density is high.
These two effects render high altitudes as favorable sites to launch dust to upper atmospheres via atmospheric escape.

It may be worth noting that intense atmospheric escape tends to yield large dust sizes at high altitudes when the particle settling takes place.
For the case of $\tau_{\rm loss}=0.1~{\rm Gyr}$ (middle and right columns of Figure \ref{fig:dust_distribution}), the dust can grow to the size of ${\sim}1~{\rm \mu m}$ at $\sim{10}^{-5}~{\rm bar}$, which is quite larger than the typical sizes ($\la0.1~{\rm \mu m}$) of photochemical hazes at the upper atmospheres of static atmospheres \citep[e.g.,][]{Kawashima&Ikoma19,Ohno&Kawashima20}.
The phenomena arise because the outflow inhibits particle settling until the particle size exceeds the threshold size. 
This was also found by \citet[][]{Gao&Zhang20} in their ``transition haze regime''.
The size enhancement can be remarkable for short $\tau_{\rm loss}$.
The effect might result in a preferential generation of flattened transmission spectra for planets with atmospheric escape, though it depends on whether the particle growth is faster than the outflow transport.

%%%%%%%%%%%%%%%%%%%%%%%%%%%%%%%%%%
\subsection{Dust Abundances in Escaping Atmospheres}\label{sec:result_abundance}
Dust mass distributions are characterized by distinct behaviors in the upper and lower regions of escaping atmospheres.
Figure \ref{fig:density_distribution} shows the vertical distribution of the dust mass mixing ratio for various $P_{\rm 0}$, $f_{\rm dust}$, and $\tau_{\rm loss}$, where the mixing ratio is given by
\begin{equation}
    w_{\rm d}= \frac{\int_{\rm 0}^{\infty}mn(m)dm}{\rho_{\rm g}}.
\end{equation}
For $P{\gg}P_{\rm 0}$, the mixing ratio decreases with decreasing the altitudes because the particle sizes increase with decreasing the altitudes, leading to enhanced settling velocities.
This trend is the same as that seen in photochemical haze models \citep[e.g.,][]{Ohno&Kawashima20,Steinrueck+20}.
On the other hand, for $P{\ll}P_{\rm 0}$, the mixing ratio is nearly invariant with altitude, as assumed in the hydrodynamic models of atmospheric escape \citep{Wang&Dai18,Wang&Dai19,Mitani+20}.

While the dust abundance is nearly constant at $P{\ll}P_{\rm 0}$, particle growth significantly affects the absolute abundances in the escaping atmospheres.
In the case of $f_{\rm dust}={10}^{-4}$ and $\tau_{\rm loss}=0.1$ Gyr (left panel of Figure \ref{fig:density_distribution}), for example, the mixing ratio is almost coincident with $f_{\rm dust}$ at upper atmospheres when the dust production pressure is $P_{\rm 0}<{10}^{-4}~{\rm bar}$.
This implies that the outflow transports almost all dust without loss through gravitational settling.
We note that the mixing ratio can be slightly higher than $f_{\rm dust}$ because the upward transport velocity is slightly slower than $v_{\rm g}$ owing to a non-zero settling velocity of dust.
On the other hand, when the production pressure is higher, the dust abundance at the upper atmospheres decreases with increasing $P_{\rm 0}$.
This is due to the fact that efficient collision growth induces the preferential dust settling before dust is transported by the outflow, as seen in Figure \ref{fig:dust_distribution}.
The abundance eventually becomes lower than $f_{\rm dust}$ by ${\sim}3$ orders of magnitude when the production pressure is $P_{\rm 0}={10}^{-2}~{\rm bar}$.

The impact of particle growth is significant, especially when the dust production rate is high.
In the case of $P_{\rm 0}={10}^{-4}~{\rm bar}$ and $\tau_{\rm loss}=0.1$ Gyr (middle panel of Figure \ref{fig:density_distribution}), the dust abundance in upper atmospheres is nearly coincident with $f_{\rm dust}$ for $f_{\rm dust}<{10}^{-4}$, whereas the abundance is significantly lower than $f_{\rm dust}$ for $f_{\rm dust}>{10}^{-4}$. 
The latter trend originates from the collision growth that is enhanced by a high dust abundance, leading to efficient gravitational settling.
Therefore, while high dust abundance is favored to increase the opacity of a dusty outflow, there is a dilemma in that the high abundance enhances the gravitational settling and hinders the upward transport of dust particles.

The outflow hardly transports the dust when the atmospheric escape is weak.
The right panel of Figure \ref{fig:density_distribution} shows the dust profile for different $\tau_{\rm loss}$.
While the dust abundance in the upper atmospheres is close to $f_{\rm dust}$ for $\tau_{\rm loss}=0.1~{\rm Gyr}$, the abundance is lower by several orders of magnitude for $\tau_{\rm loss}\ge 1~{\rm Gyr}$.
This result demonstrates that it is more challenging to blow up dust on planets with long $\tau_{\rm loss}$.

%\subsection{Dust Abundance in Escaping Atmospheres}
%%%%%%%%%%%%%%%%%%%%%%%%%%%%%%%%%%%
\begin{figure}[t]
\centering
\includegraphics[clip, width=\hsize]{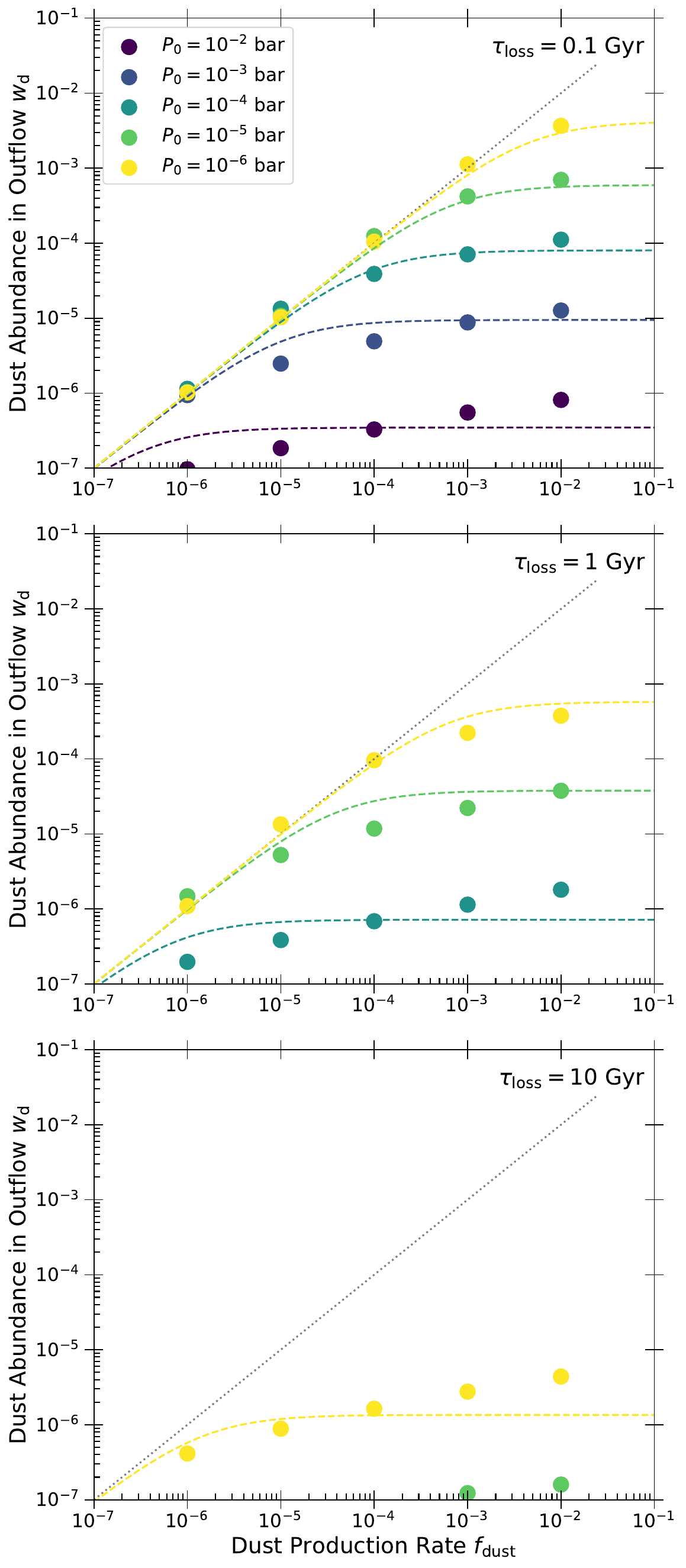}
\caption{Dust abundance in the escaping atmosphere as a function of $f_{\rm dust}$. Different colored points denote the dust mass mixing ratio measured at ${10}^{-9}~{\rm bar}$ for different dust production altitudes $P_{\rm 0}$. From top to bottom, each panel shows the result for $\tau_{\rm loss}=0.1$, $1$, and $10~{\rm Gyr}$, respectively. The gray dotted line denotes a $w_{\rm d}=f_{\rm dust}$ relation. The colored dashed lines denote the dust abundance predicted by Equation \eqref{eq:wd_main}.}
\label{fig:wd_summary}
\end{figure}
%%%%%%%%%%%%%%%%%%%%%%%%%%%%%%%%%%%

Figure \ref{fig:wd_summary} summarizes how the dust abundances in upper atmospheres vary with relevant parameters.
Although \citet{Wang&Dai19} assumed the dust abundance of $w_{\rm d}{\sim}{10}^{-2}$, such a high abundance can be achieved only under restricted conditions, namely short mass-loss timescales $\tau_{\rm loss}=0.1~{\rm Gyr}$ and high dust production altitudes ($P_{\rm 0}={10}^{-6}~{\rm bar}$).
The production altitude is particularly important: the dust abundance is only $w_{\rm d}{\sim}{10}^{-6}$ even in the case of $\tau_{\rm loss}=0.1~{\rm Gyr}$ and $f_{\rm dust}={10}^{-2}$ when the dust are formed at $P_{\rm 0}={10}^{-2}~{\rm bar}$.
Thus, we suggest that dust must be formed at sufficiently high altitudes, say $\la{10}^{-5}~{\rm bar}$, to launch the dusty outflow with high dust abundances.
This result provides constraints on what dust formation processes are likely responsible for the dusty outflow, which will be discussed in Section \ref{sec:dis_dust_nature}.

There is a general trend that the dust abundance almost coincides with $f_{\rm dust}$ for low $f_{\rm dust}$ and plateaus at high $f_{\rm dust}$.
To better understand this trend, we construct an analytic theory that predicts the dust abundances in escaping atmospheres. 
The theory calculates the dust abundance in the outflow by taking into account collision growth assuming a single particle size and zero settling velocity.
The dust abundance in the upper escaping atmosphere can be estimated as (see Appendix \ref{sec:anal_derivation0} for the derivation)
\begin{equation}\label{eq:wd_main}
    w_{\rm d}\approx \frac{f_{\rm dust}}{1+\chi}.
\end{equation}
As seen in Equation \eqref{eq:wd_main}, the dust abundance is controlled by a single dimensionless parameter $\chi$ that is the ratio of the upward transport to the collision growth timescales for $w_{\rm d}=f_{\rm dust}$ at the dust production altitude given by (Equation \ref{eq:chi_appro0} for an original definition)
\begin{eqnarray}\label{eq:chi_main}
     \chi \approx 0.23\frac{f_{\rm dust}}{\mathcal{M}}\left( 1+\frac{256m_{\rm g}}{\pi m_{\rm cri}}\mathcal{M}^{-2}\right)^{1/2},
\end{eqnarray}
where $\mathcal{M}\equiv v_{\rm g}(P_{\rm 0})/c_{\rm s}$ is the Mach number at the dust production pressure, and $m_{\rm cri}=4\pi a_{\rm cri}^3\rho_{\rm p}/3$ is the mass of dust with the threshold size of Equation \eqref{eq:a_cri}.
The upward transport is faster than the collision growth for $\chi<1$, and vice versa for $\chi>1$.
Equation \eqref{eq:wd_main} has asymptotic forms of
\begin{eqnarray}\label{eq:wd_limit}
  w_{\rm d} \approx \left\{ \begin{array}{ll}
    {\displaystyle f_{\rm dust} }& (\chi \ll 1) \\
   {\displaystyle 4.35 \mathcal{M} \left( 1+\frac{256m_{\rm g}}{\pi m_{\rm cri}}\mathcal{M}^{-2}\right)^{-1/2}} & (\chi \gg 1)
  \end{array} \right..
\end{eqnarray}
In the limit of fast upward transport ($\chi\ll1$), the abundance approaches to $f_{\rm dust}$.
In the opposite limit of efficient particle growth ($\chi\gg1$), the abundance is regulated to be lower than the Mach number of the dust production altitude.
This regulation comes from the timescale of settling-driven collisions, {the so-called the coalescence \citep{Rossow78}}, given by \citep[Equation (29) of][]{Ohno&Okuzumi18}
\begin{equation}\label{eq:tau_coal}
    \tau_{\rm coal}\sim\frac{r^2c_{\rm s}}{GM_{\rm p}w_{\rm d}}=\frac{\mathcal{M}}{w_{\rm d}}\tau_{\rm tran},
\end{equation}
where $\tau_{\rm tran}{\equiv}H/v_{\rm g}$ is the upward transport timescale, and $H=r^2c_{\rm s}^2/GM_{\rm p}$ is the pressure scale height.
Equation \eqref{eq:tau_coal} indicates that the settling-driven collision always dominates over the upward transport as long as $w_{\rm d}{>}\mathcal{M}$.
The only way to halt the collision growth is to reduce the dust abundance until $w_{\rm d}{\la}\mathcal{M}$, which can be caused by gravitational settling.
This timescale argument explains why the abundance is regulated to $\la\mathcal{M}$.
Although little deviations attributed to size distributions and a nonzero settling velocity appear in high $P_{\rm 0}$ and $f_{\rm dust}$ cases, Equation \eqref{eq:wd_main} reasonably reproduces the numerical results within a factor of $\sim3$ in Figure \ref{fig:wd_summary}.
Because of analytic nature, the theory can be easily utilized in hydrodynamical models \citep[e.g.,][]{Wang&Dai18,Wang&Dai19,Mitani+20} as well as in thermal evolution models coupled with the escape models \citep[e.g.,][]{Lopez+12,Lopez&Fortney13,Owen&Wu13,Kurosaki+14,Kurokawa&Nakamoto14,Chen&Rogers16,Kubyshkina+20} to evaluate the dust abundances in upper escaping atmospheres.

%%%%%%%%%%%%%%%%%%%%%%%%%%%%%%%%%%%
\section{Observational Implications}\label{sec:observation}
\subsection{Impacts on Transit Radius}\label{sec:impact_radius}
%Since it is suggested that dusty outflow can enhance the observable planetary radius and explain the large radii of super-puffs \citep{Wang&Dai19}, 
We investigate how much the dusty outflow can enhance the observable radius when the particle growth is taken into account.
We calculate the effective transit radius as
\begin{equation}\label{eq:R_eff}
    R_{\rm eff}^2=R_{\rm 0}^2+2 \int_{\rm R_{\rm 0}}^{R_{\rm H}}[1-\exp{(-\tau_{\rm s})}]rdr,
\end{equation}
where $\tau_{\rm s}$ is the line-of-sight chord optical depth, $R_{\rm 0}$ is the reference radius \citep[e.g.,][]{Heng19}, $R_{\rm H}=a_{\rm orb}(M_{\rm p}/3M_{\rm s})^{1/3}$ is the Hill radius, and $M_{\rm s}$ is the stellar mass.
We set $R_{\rm 0}$ to the lower boundary of computation domain, corresponding to $P=10~{\rm bar}$, and assume $a_{\rm orb}=0.1~{\rm AU}$ and solar mass for evaluating $R_{\rm H}$.
We select the Hill radius as an upper end of the integration because spherical symmetry assumed in our model is no longer valid for $r>R_{\rm H}$.
We also tested the upper end of $r=R_{\rm s}$, where the stellar radius $R_{\rm s}$ is set to the sun radius, and confirmed that the results are almost unchanged.  
The atmospheric opacity is calculated as
\begin{equation}
    \kappa=\kappa_{\rm gas}+\int_{\rm 0}^{\infty}\pi a^2 Q_{\rm ext}(a,\lambda)n(m)dm,
\end{equation}
where $\kappa_{\rm gas}$ is the gas opacity, $\lambda$ is the wavelength, and $Q_{\rm ext}$ is the extinction efficiency.
The wavelength and gas opacity are set to $1~{\rm \mu m}$ and $\kappa_{\rm gas}=3\times{10}^{-3}~{\rm {cm}^{2}~g^{-1}}$ in this subsection, respectively.
We examine the ratio of the transit radius with the dusty outflow to that without to evaluate the magnitude of the radius enhancement.
For comparisons, we also calculate the radius for the growth-free dusty outflow by assuming $w_{\rm d}=f_{\rm dust}$ and $a=0.001~{\rm {\mu}m}$ in entire atmospheres.

\begin{figure}[t]
\centering
\includegraphics[clip, width=\hsize]{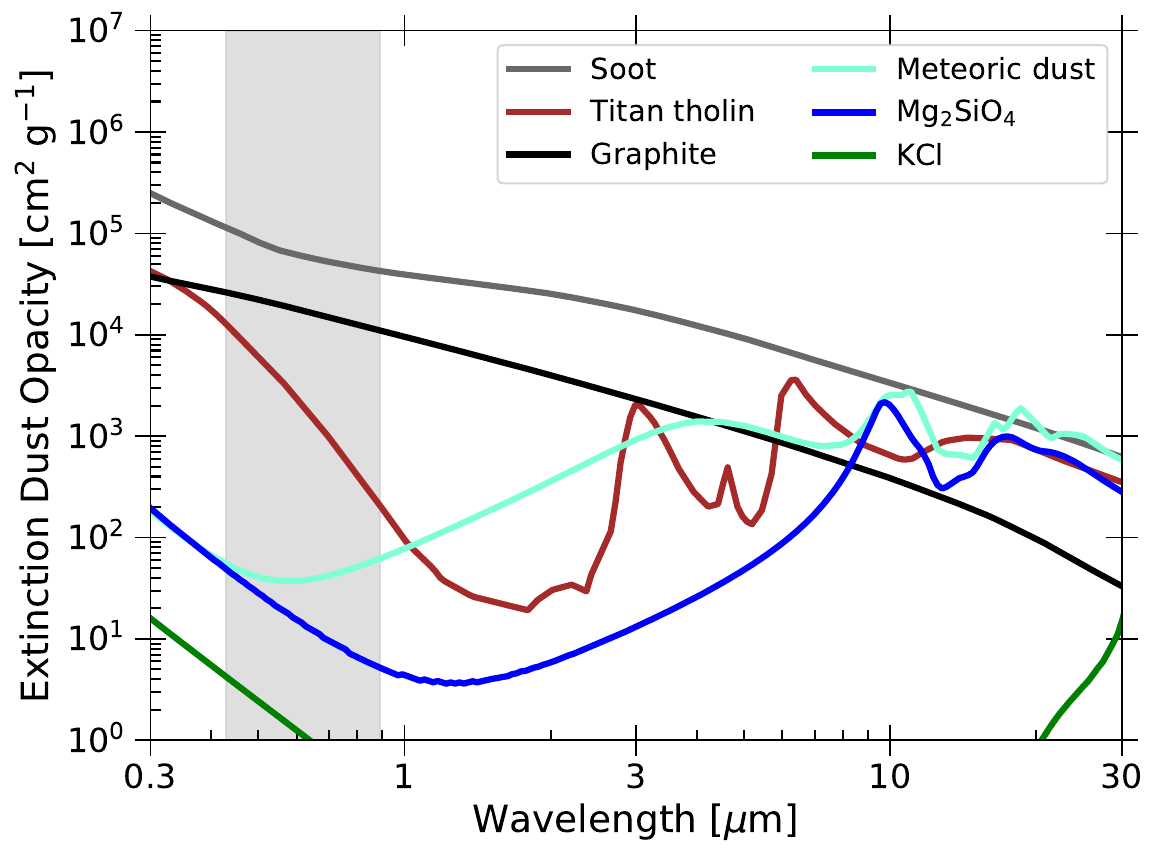}
\caption{Extinction opacity of some representative atmospheric dust. Particle radius is set to $0.01~{\rm \mu m}$ so that dust can be entrained in the escaping atmosphere (Equation \ref{eq:a_cri}). We select the optical constants of soot \citep{Lavvas&Koskinen17}, Titan tholin \citep{Khare+84}, graphite \citep{Draine03}, meteoric dust \citep{Fenn+85}, Mg$_2$SiO$_4$ \citep{Jager+03}, and KCl \citep{Palik85}.
The material density is set to $1.00~{\rm g~{cm}^{-3}}$ for soot and Titan tholin, $2.00~{\rm g~{cm}^{-3}}$ for graphite and meteoric dust,  $3.21~{\rm g~{cm}^{-3}}$ for Mg$_2$SiO$_4$, and $1.98~{\rm g~{cm}^{-3}}$ for KCl.
The gray shaded region denotes the wavelength bandpass of the  Kepler space telescope.
}
\label{fig:Mie_opacity}
\end{figure}

We compute the extinction efficiency of dust particles using the Mie theory code of \citet{Bohren&Huffman83} with the refractive index of complex refractory carbon (soot) compiled by \citet{Lavvas&Koskinen17}, which has been used as a spectral analog of photochemical hazes formed in hot exoplanetary atmospheres.
We note that the radius enhancement estimated here is presumably an upper limit because soot opacity is considerably higher than other candidates of exoplanetary aerosols' opacity.
Figure \ref{fig:Mie_opacity} shows the opacity of several exoplanetary aerosol analogs.
While the soot opacity is similar to the graphite opacity assumed in \citet{Wang&Dai19}, this opacity is considerably higher than the opacity of Mg$_2$SiO$_4$, KCl, and Titan tholin.
Because the outflow can transport only tiny particles, and the scattering efficiency steeply decreases with decreasing particle sizes \citep[$Q_{\rm sca}{\propto}a^4$ at $2\pi a\ll\lambda$,][]{Bohren&Huffman83}, absorbing materials like soot act to increase the atmospheric opacity much more efficiently than the scattering materials like KCl do. 
Future retrieval studies may constrain optical constants relevant to exoplanetary aerosols from an observational perspective \citep{Taylor+20}.

\begin{figure}[t]
\centering
\includegraphics[clip, width=\hsize]{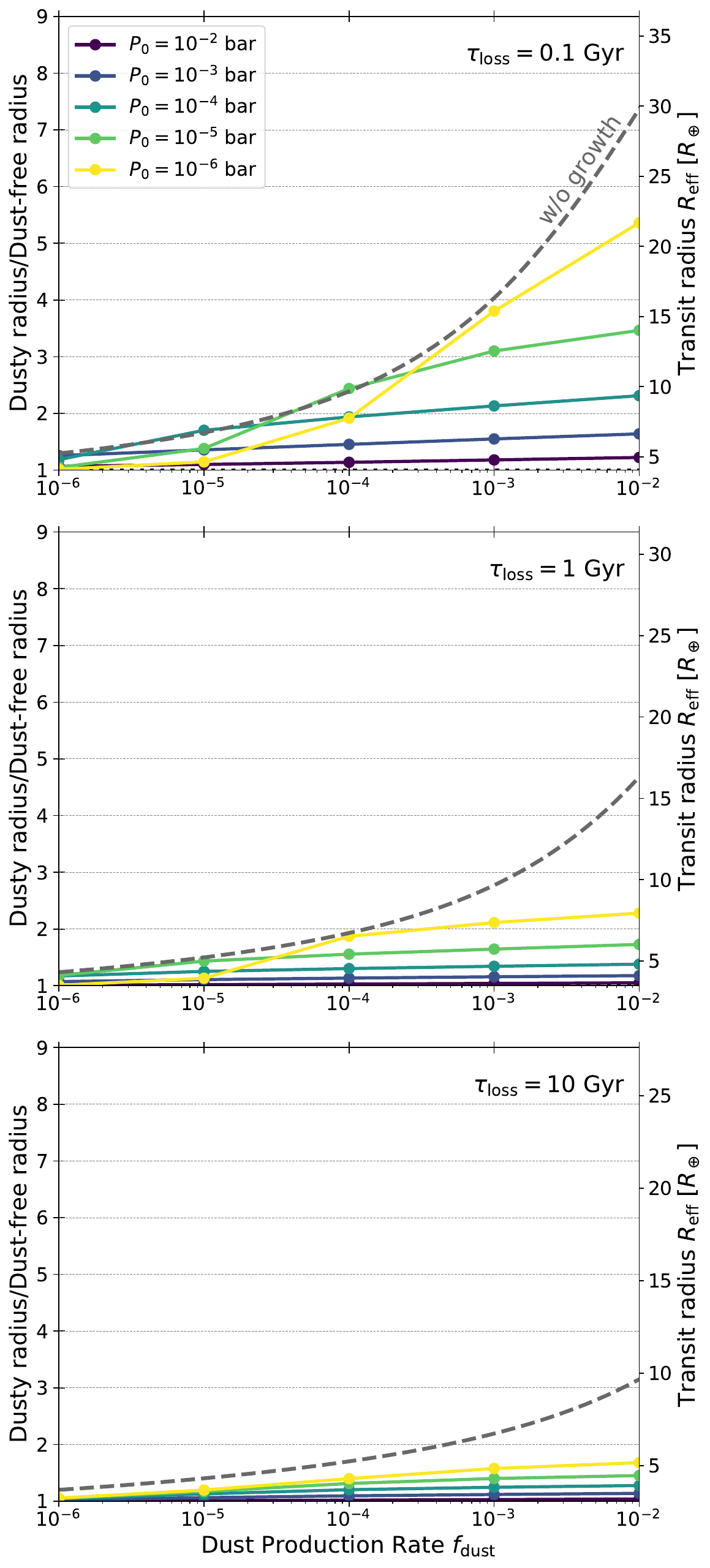}
\caption{Effective transit radii of a Kepler-51b like super-puff. The vertical axis is the radius with dusty outflow normalized by the radius with a dust-free atmosphere, and the horizontal axis is the dust production rate $f_{\rm dust}$. Different colored lines show the result for different $P_{\rm 0}$. From top to bottom, each figure shows the result for $\tau_{\rm loss}=0.1$, $1$, and $10~{\rm Gyr}$, respectively. The gray dashed lines show the radii with vertically constant dust sizes and abundances ($a=0.001~{\rm {\mu}m}$ and $w_{\rm d}=f_{\rm dust}$), representing the results for growth-free cases.
}
\label{fig:Reff_summary}
\end{figure}

The dusty outflow could enhance the transit radius significantly if the particle growth is neglected.
Figure \ref{fig:Reff_summary} shows the ratio of the transit radii with the dusty outflow to that without.
The radius enhancement becomes drastic as the dust production rate increases.
For example, the radius is larger than the dust-free case only by $\la10\%$ for $f_{\rm dust}={10}^{-6}$, while the radius is enhanced by a factor of $\sim3$--$7$ for $f_{\rm dust}={10}^{-2}$, depending on the mass-loss timescale.
This is consistent with \citet{Wang&Dai19} who showed that the transit radius is enhanced by a factor of $\sim3$ from the dust-free case for $w_{\rm d}\sim2$--$5\times{10}^{-2}$ when the mass-loss timescale is $\sim3$--$5~{\rm Gyr}$.

%%%%%%%%%%%%%%%%%%%%%%%%%%%%%%%%%%%
%%%%%%%%%%%%%%%%%%%%%%%%%%%%%%%%%%%

%%%%%%%%%%%%%%%%%%%%%%%%%%%%%%%%%%%
%%%%%%%%%%%%%%%%%%%%%%%%%%%%%%%%%%%

The particle growth significantly reduces the radius enhancement caused by the dusty outflow when the dust is formed at lower atmospheres.
In the case of $P_{\rm 0}={10}^{-2}~{\rm bar}$, for example, the outflow can increase the radius only by ${\sim}10\%$ even for a high dust production rate of $f_{\rm dust}={10}^{-2}$ and a short mass-loss timescale of $\tau_{\rm loss}=0.1~{\rm Gyr}$.
The dusty outflow tends to cause a more considerable radius enhancement as the dust production altitude is shifted to higher altitudes.
For example, a dusty outflow could enhance the transit radius by more than $\sim50\%$ for $\tau_{\rm loss}=1~{\rm Gyr}$ if dust particles are formed at $P_{\rm 0}\la {10}^{-5}~{\rm bar}$.
This radius enhancement is a consequence of efficient outflow transport that increases the dust abundance in upper atmospheres, as shown in Section \ref{sec:result_abundance}.
These results strongly suggest that only atmospheric dust formed in upper atmospheres can be responsible for enhancing the observable planetary radius.

The above explanation does not mean that an unlimited high production altitude is favored to enhance the observable radius, as very high production altitudes sometimes cause a small transit radius.
This can be seen in $\tau_{\rm loss}=1$ and $0.1~{\rm Gyr}$ cases where the transit radius for $P_{\rm 0}={10}^{-6}~{\rm bar}$ is smaller than the radii for $P_{\rm 0}={10}^{-5}$--${10}^{-3}~{\rm bar}$ at $f_{\rm dust}<{10}^{-4}$.
The trend originates from the absence of dust below the production altitude when the outflow completely blows up dust to upper atmospheres (see Figure \ref{fig:density_distribution}). 
If the dust production rate is so low that the dusty outflow is optically thin in this circumstance, the transit radius is mainly determined by the opacity of lower dust-free atmospheres.
This is why the transit radius is close to that for dust-free atmospheres in very low $f_{\rm dust}$ and $P_{\rm 0}$ cases. 
The upper limit on the dust production altitude may be evaluated as follows.
When the particle growth is neglected, the slant optical depth of the isothermal atmosphere can be approximated by
\begin{equation}\label{eq:tau_slant}
    \tau_{\rm s} \approx \rho_{\rm g}f_{\rm dust}\kappa_{\rm dust} \sqrt{2\pi rH},
\end{equation}
where $\kappa_{\rm dust}$ is the dust opacity, and we crudely assume a constant planetary gravity.
The dusty outflow is optically thin when $\tau_{\rm s}<1$ at $\rho_{\rm g}=P_{\rm 0}/c_{\rm s}^2$.
Solving $\tau_{\rm s}=1$ using the scale height of $H=r^2c_{\rm s}^2/GM_{\rm p}$ and $r{\sim}R_{\rm p}$, we obtain the minimum production pressure below which the dusty outflow is optically thin as
\begin{eqnarray}
    P_{\rm min} &\sim& \frac{c_{\rm s}}{f_{\rm dust}\kappa_{\rm dust}} \sqrt{\frac{2\rho_{\rm p} G}{3}}\\
    \nonumber
    & \sim & 2\times{10}^{-6}~{\rm bar}~\left( \frac{f_{\rm dust}\kappa_{\rm dust}}{10~{\rm {cm}^2~g^{-1}}}\right)^{-1} \left( \frac{c_{\rm s}}{1~{\rm {km}~s^{-1}}}\right)\left( \frac{\rho_{\rm p}}{1~{\rm {g}~{cm}^{-3}}}\right)^{1/2},
\end{eqnarray}
where $\rho_{\rm p}$ is the planetary bulk density.
Note that this condition applies only when the particle growth is negligible (i.e., $\chi\ll1$).
Thus, dust needs to be formed at as high an altitude as possible while satisfying $P_{\rm 0}>P_{\rm min}$ in order to maximize the enhancement of the observable radius.

How much the dusty outflow can enhance the transit radius also varies with the mass-loss timescales.
While the intense atmospheric escape does help to blow up dust, it is worth noting that a shorter $\tau_{\rm loss}$ causes a larger radius enhancement even for growth-free cases where all dust is transported to upper atmospheres regardless of $\tau_{\rm loss}$.
In the growth-free case at $f_{\rm dust}={10}^{-2}$, for example, the outflow enhances the transit radius by a factor of $\sim3$ for $\tau_{\rm loss}=10~{\rm Gyr}$, whereas the radius is enhanced by a factor of $\sim7$ for $\tau_{\rm loss}=0.1~{\rm Gyr}$.
This stems from the fact that, in the context of an isothermal atmosphere, a short mass-loss timescale is caused by a high atmospheric density at a sonic point, which implies that isobaric planes are elevated to high altitudes.
The atmospheric density at the transit radius almost solely depends on dust opacity $w_{\rm d}\kappa_{\rm dust}$ (see Equation \ref{eq:tau_slant}), and thus the transit radius becomes larger as the isobaric planes are elevated to higher altitudes.
The short mass-loss timescale is actually not the direct cause of the radius enhancement but the inevitable outcome of the enhancement. 
We return to this topic in Section \ref{sec:MR}.

%%%%%%%%%%%%%%%%%%%%%%%%%%%%%%%%%%%%
\subsection{Inferences on Dust Formation Processes}\label{sec:dis_dust_nature}
Our results suggest that atmospheric dust could enhance the observable radius only when formed at high altitudes.
In other words, the dusty outflow scenario of \citet{Wang&Dai19} is viable only when the dust is produced in upper atmospheres.
This supports the idea that photochemical hazes may explain the large radii of super-puffs \citep[][]{Gao&Zhang20} because they are typically formed at  $\la{10}^{-5}~{\rm bar}$.
Recent laboratory studies have confirmed the production of photochemical hazes in relatively hot environments relevant to exoplanetary atmospheres \citep[e.g.,][]{Horst+18,He+18,Fleury+19,Moran+20}.
The haze hypothesis is also compatible with the typical temperature of super-puffs ($<1000~{\rm K}$) at which CH$_4$ is the main carbon reservoir and promotes haze formation \citep{Morley+15,Kawashima&Ikoma19,Gao+20}.
Although condensation clouds of alkali metals, such as KCl, are expected to form in the same temperature regimes \citep[e.g.,][]{Miller-Ricci+12,Morley+13,Mbarek&Kempton16,Lee+18}, they are unlikely to be responsible for the radius enhancement for super-puffs because of low formation altitudes (typically $P\ga{10}^{-2}~{\rm bar}$ in pressure).
This is also qualitatively consistent with previous microphysical models that indicated the difficulty of condensation clouds to ascend to upper atmospheres \citep{Ohno&Okuzumi18,Gao&Benneke18}.
We also point out that KCl clouds have a very low visible opacity, as shown in Figure \ref{fig:Mie_opacity}, which further renders the condensation clouds a nonviable candidate to enhance the observable radius.

The ablation of meteoroids infalling to a planet may also produce dust at high altitudes.
In the Earth atmosphere, the incoming meteoroids are mainly evaporated at the altitude of $80$--$100~{\rm km}$ ($<{10}^{-5}~{\rm bar}$ in pressure) and recondensed into tiny smoke particles \citep[e.g.,][]{Hunten+80,Moliza-Cuberos+08,Plane+18}.
The presence of such meteoric dust at high altitudes is universal in other system objects \citep[e.g.,][]{Moses92,McAuliffe&Christou06,Moses&Poppe17}. 
\citet{Lavvas&Koskinen17} discussed that the meteoric dust may also emerge at $\la{10}^{-6}~{\rm bar}$ in hot Jupiter atmospheres.
It is currently difficult to quantify how much meteoric dust is injected into exoplanetary atmospheres and affects the observable radius.
However, we suggest that it may be possible to investigate whether the meteoric dust plays an important role or not by searching for silicate features in atmospheric transmission spectra, as discussed in the next section.

%%%%%%%%%%%%%%%%%%%%%%%%%%%%%%%%%%%%%
\subsection{Transmission Spectra of the Dusty Outflow}\label{sec:spectrum}
%%%%%%%%%%%%%%%%%%%%%%%%%%%%%%%%%%%
\begin{figure*}[t]
\centering
\includegraphics[clip, width=\hsize]{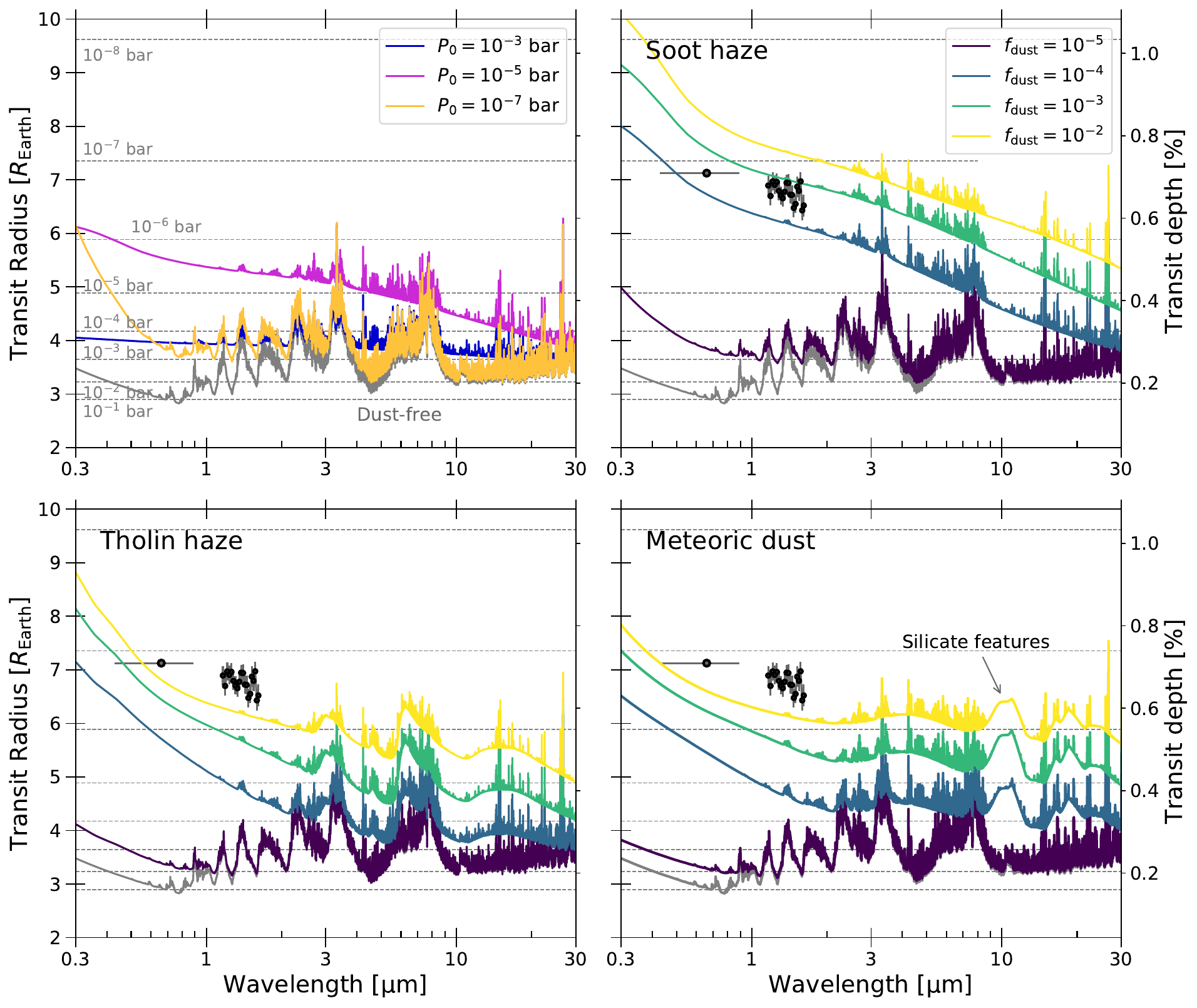}
\caption{Synthetic transmission spectra of Kepler-51b like super-puffs with dusty outflow. The left top panel shows the spectra with $f_{\rm dust}={10}^{-4}$ for different dust production height, while other panels show the spectra with $P_{\rm 0}={10}^{-6}~{\rm bar}$ for different dust production rates. 
We use the refractive indices of soot for the top two panels, tholin haze for the left bottom panel, and meteoric dust for the right bottom panel, respectively.
The gray spectra show the spectra for dust-free atmosphere, the gray dotted lines show the pressure levels from ${10}^{-1}$ to ${10}^{-8}~{\rm bar}$, and the black dots denote the spectra of Kepler-51b observed by Kepler \citep{Masuda14} and HST-WFC3 \citep{Libby-Roberts+20}, where we use the latest stellar radius of Kepler-51 \citep[$0.881~{R_{\rm sun}}$,][]{Libby-Roberts+20}.
The right vertical axes exhibit the corresponding transit depth.
We assume an atmospheric structure with $\tau_{\rm loss}=1~{\rm Gyr}$, and the all spectra are binned down to $\lambda/\Delta \lambda=1000$.
}
\label{fig:puff_spectrum}
\end{figure*}
%%%%%%%%%%%%%%%%%%%%%%%%%%%%%%%%%%%

It is intriguing to see how the dusty outflow affects the observable spectra, as several super-puffs exhibit featureless transmission spectra \citep{Libby-Roberts+20,Chachan+20}. 
Here, we compute the synthetic spectra with a dusty outflow using a model of \citet{Ohno+20a} with some updates.
The calculation procedure is to solve Equation \eqref{eq:R_eff} at multiple wavelengths by taking into account gas and dust opacity.
We include the gas absorption of H$_2$O, CH$_4$, CO, CO$_2$, NH$_3$, and HCN, where the line opacity for each molecule is obtained from the Dace Opacity database \citep{Grimm+21}\footnote{\href{https://dace.unige.ch/dashboard/index.html}{https://dace.unige.ch/dashboard/index.html}}. 
We also include the collision-induced absorption of H$_2$--H$_2$ and H$_2$--He from HITRAN \citep{Richard+12,Karman+19} as well as the Rayleigh scattering by H$_2$ \citep{Dalgarno&Williams62} and He \citep{Chan&Dalgarno65}.
For the dust opacity, we test three representative refractive indices: Titan tholin \citep{Khare+84} and soot \citep{Lavvas&Koskinen17} as spectral analogs of photochemical hazes, and meteoric dust \citep{Fenn+85}.
The abundance of each molecule is calculated by the publicly available code GGchem \citep{Woitke+18} under the assumption of thermochemical equilibrium in a solar composition atmosphere at $T=500~{\rm K}$.
Although the atmospheres of warm exoplanets are potentially depleted in CH$_4$ due to several mechanisms, such as disequilibrium chemistry and cloud radiative feedback \citep{Molaverdikhani+20,Fortney+21}, which was also suggested by observations \citep{Stevenson+10,Kreidberg+18,Benneke+19}, we neglect the possible CH$_4$ depletion as it does not affect the conclusions of this paper.  

In general, the dusty outflow obscures the absorption features of gas molecules, and the degree to which it does highly depends on the dust production altitudes.
The left top panel of Figure \ref{fig:puff_spectrum} shows the transmission spectra for different dust production altitudes, where we set $\tau_{\rm loss}=1~{\rm Gyr}$ and $f_{\rm dust}={10}^{-4}$. 
In the visible to near-infrared wavelengths, the dust-free atmosphere yields the transit radius at $\sim3~R_{\rm \oplus}$, which is corresponding to $P\sim{10}^{-1}$--${10}^{-3}~{\rm bar}$. 
When the dust with soot optical constant is included, the dusty outflow moderately obscures the spectral features and increases the transit radius to $\sim4R_{\rm \oplus}$ for $P={10}^{-3}~{\rm bar}$.
The visible spectrum is mostly flat because particle sizes tend to be large in the escaping atmospheres when the settling dominates over the outflow (Section \ref{sec:result1}). 
The dust abundance in the outflow increases with increasing the production altitude, and hence the dusty outflow mostly obscures the spectral features and increases the radius to $\sim6R_{\rm \oplus}$ for $P_{\rm 0}={10}^{-5}~{\rm bar}$.
The spectrum becomes a considerably different shape characterized by a steep spectral slope and rich gas features when the dust production altitude is too high, as seen in the case of $P_{\rm 0}={10}^{-7}~{\rm bar}$.
The rich gas features originate from that the optically thin dusty outflow.
The steep slope at short wavelengths is attributed to the vertical opacity gradient, in which atmospheric opacity is higher at higher altitudes \citep{Ohno&Kawashima20}. 
Since the outflow blows up most of the dust for very low $P_{\rm 0}$, atmospheric opacity steeply decreases at $P>P_{\rm 0}$ owing to the absence of dust, resulting in the steep vertical opacity gradient.
The phenomenon presented here is an analogy of the so-called ''cloud-base effect`` \citep{Vahidinia+14} that causes a steep drop in the transit depth when the opaque pressure level crosses the cloud-base pressure.

Transmission spectra of super-puffs with dusty outflows tend to exhibit spectral slopes when the appreciable radius enhancement is achieved.
This trend may be inevitable because the outflow can only contain submicron dust.
The large pressure scale height in upper atmospheres further steepens the spectral slope.
Thus, we predict that radii of super-puffs appreciably decrease with increasing the wavelength if the dusty outflow is indeed the cause of their large radii.
The significant decrease of the planetary radius at long wavelengths was also suggested by \citet{Gao&Zhang20}.

Since our hypothetical super-puff has a mass similar to that of Kepler-51b, it is worth investigating whether the model can explain the existing observations.
Figure \ref{fig:puff_spectrum} compares the observed transit radius of Kepler-51b \citep{Masuda14,Libby-Roberts+20} with the synthetic spectra for different dust production rate and optical constants under the assumption of $P_{\rm 0}={10}^{-6}~{\rm bar}$ and $\tau_{\rm loss}=1~{\rm Gyr}$.
We found that the synthetic spectrum with soot optical constant well matches the observed transit radii when the production rate is $f_{\rm dust}\sim3\times{10}^{-4}$.
The dusty outflow mostly eliminates spectral features in visible to near-infrared wavelength, consistent with the observations by \citet{Libby-Roberts+20}.
Our result confirms the finding of \citet{Gao&Zhang20} who also succeeded in explaining the observations by soot hazes.
Although visible and NIR spectra are mostly featureless, some absorption lines begin to appear at wavelengths longer than $\sim2~{\rm \mu m}$.
This potentially suggests that observations at $\sim2~{\rm \mu m}$ may be able to detect gas species even for super-puffs with nearly flat near-infrared spectra.
Such long wavelengths will be accessible for upcoming observations by NIRSpec ($0.7$--$5~{\rm \mu m}$) and MIRI ($5.0$--$14~{\rm \mu m}$) on the JWST \citep{Batalha+17}, Twinkle \citep[$0.4$--$4.5~{\rm \mu m}$,][]{Edwards+19}, and ARIEL \citep[$1.25$--$7.8~{\rm \mu m}$,][]{Tinetti+18}. 
Alternatively, high-resolution transmission spectroscopy with ground-based telescopes \citep[e.g.,][]{Birkby18,Pino+18a,Pino+18b,Molliere&Snellen19,Hood+20,Gandhi+20} would also be a promising way to detect gas species in super-puff atmospheres.

We also find that the synthetic spectra hardly explain the observations of Kepler-51b when the optical constants of Titan tholin and meteoric dust are assumed.
This stems from their less absorptive optical constants in the near-infrared, resulting in low extinction opacity for tiny dust entrained in the outflow (see Figure \ref{fig:Mie_opacity}).
Thus, aerosols made of absorptive materials, like soot and graphite, are favored to explain the observations of Kepler-51b.

Although meteoric dust has a relatively minor impact on the observed radii, it is interesting to note that it produces silicate features at $\lambda{\sim}10~{\rm \mu m}$.
The emergence of silicate features has been anticipated for silicate clouds in hot Jupiters \citep{Powell+19,Gao+20}, mineral atmospheres on magma ocean planets \citep{Ito+15,Ito+21}, and dust tails of ultra-short period disintegrating planets \citep{Bodman+18,Okuya+20}, but not expected for the temperature regimes of super-puffs. 
Though this is beyond the topic of this study, searching for silicate features in cool exoplanetary atmospheres may help to constrain how many meteoroids are infalling to the planet.

%%%%%%%%%%%%%%%%%%%%%%%%%%%%%%%%%%%%%
%\subsection{Influence on Mass-Radius Relation}\label{sec:MR}

%%%%%%%%%%%%%%%%%%%%%%%%%%%%%%%%%%%%%%%%%%%%%%%%%%%%%%%%%%%%
\section{Why are Super-puffs Uncommon?}\label{sec:MR}
If the large radii of super-puffs are caused by photochemical hazes, this poses a question: why are super-puffs uncommon?
Observations of atmospheric transmission spectra frequently reported more-or-less featureless spectra for low-mass warm ($<1000~{\rm K}$) exoplanets \citep[e.g.,][]{Kreidberg+14,Kreidberg+18,Kreidberg+20,Knutson+14b,Knutson+14a,Benneke+19,Libby-Roberts+20,Chachan+19,Chachan+20,Guo+20,Mikai-Evans+21,Guilluy+21}.
From these observations, several studies have suggested that photochemical hazes may universally exist in warm exoplanetary atmospheres \citep[e.g.,][]{Crossfield&Kreidberg17,Gao+20}.
However, only a small fraction of whole low-mass planets exhibit extremely low bulk density \citep[${\sim}15\%$,][]{Cubilios+17}, which may appear to contradict the hypothesis that haze inflates the apparent radii of super-puffs.
In what follows, we demonstrate that the universality of hazes and the rarity of super-puffs can coexist because high-altitude hazes cause significant radius enhancement only for planets near the stability limit.

%%%%%%%%%%%%%%%%%%%%%%%%%%
\begin{figure}[t]
\centering
\includegraphics[clip, width=\hsize]{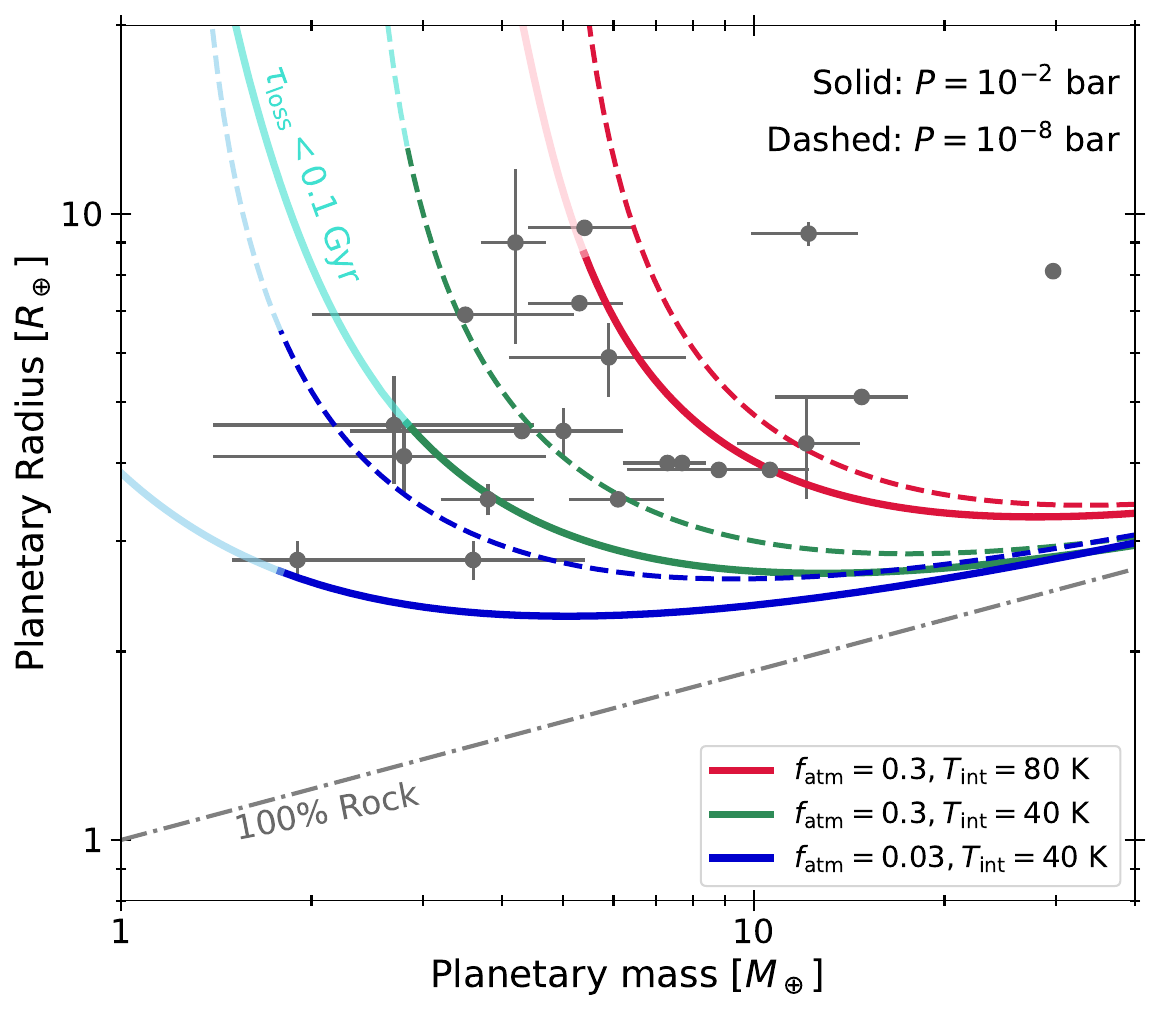}
\caption{Mass-radius relation of rocky planets surrounded by solar composition atmospheres computed by an interior structure model described in Appendix \ref{appendix:MR}. The red and green lines show the relation for intrinsic temperature of $T_{\rm int}=80$ and $40~{\rm K}$, respectively, with $f_{\rm atm}=0.3$. 
The blue line shows the relation for $T_{\rm int}=40$ and $f_{\rm atm}=0.03$. 
The solid lines show the relation defining planet radii at $P={10}^{-2}~{\rm bar}$, while the dashed lines show the radii at $P={10}^{-8}~{\rm bar}$.
The gray dash-dot line shows the core radius of Earth-like composition \citep{Zeng+19}, and the gray plots exhibit the observed mass-radius relation of super-puff candidates, listed in \citet{Chachan+20}.
The thin colored parts denote the parameter spaces for the mass-loss timescale of $\tau_{\rm loss}<0.1~{\rm Gyr}$.
}
\label{fig:MR_diagram}
\end{figure}
%%%%%%%%%%%%%%%%%%%%%%%%%%%%%%%%%%%
\begin{figure}[t]
\centering
\includegraphics[clip, width=0.98\hsize]{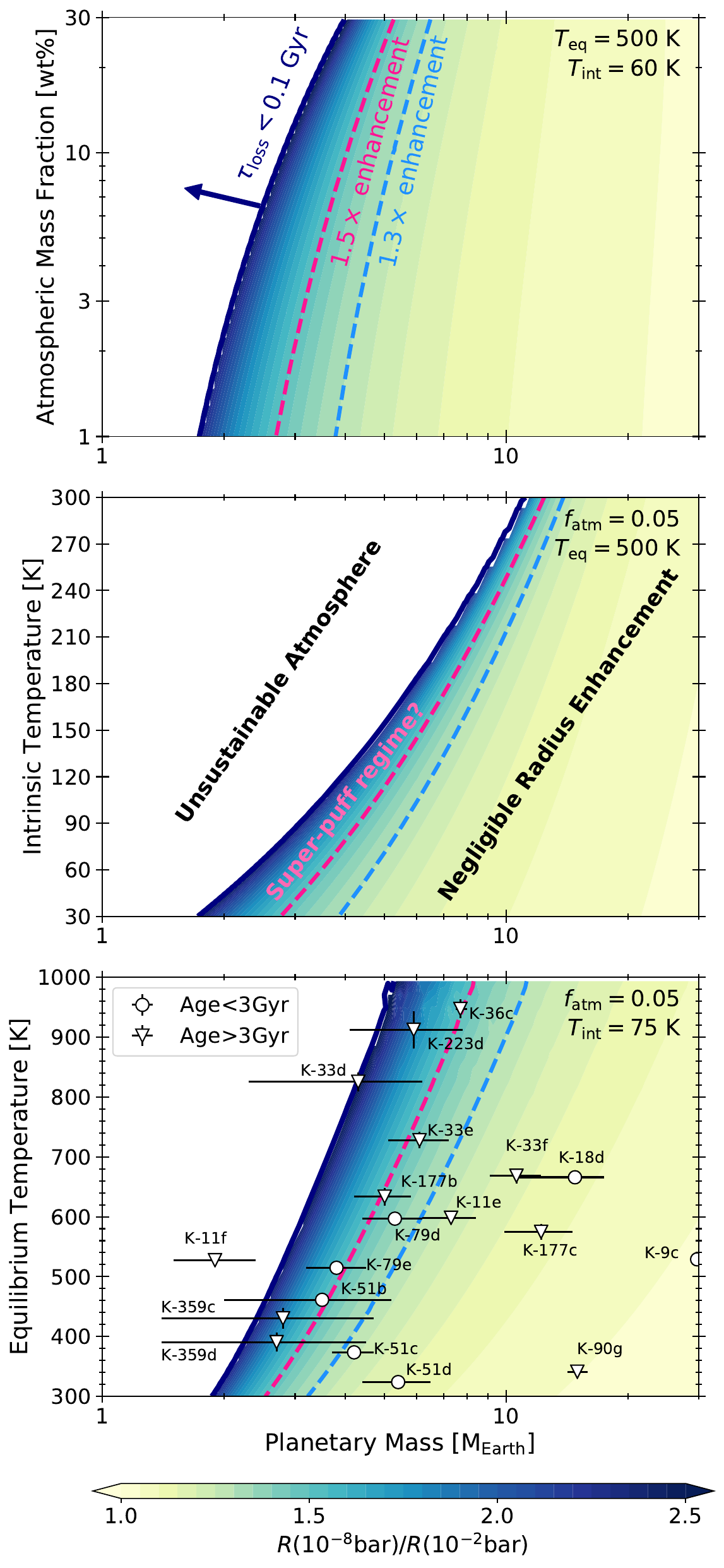}
\caption{Ratio of planetary radii at $P={10}^{-8}~{\rm bar}$ to $P={10}^{-2}~{\rm bar}$ (colorscale). The horizontal axis is planetary mass, and the vertical axis is an atmospheric mass fraction, planetary intrinsic temperature, and equilibrium temperature from top to bottom, respectively.
The red and blue dashed lines show the contour of $R({10}^{-8}~{\rm bar})/R({10}^{-2}~{\rm bar})=1.5$ and $1.3$, respectively, and the white {  unsustainable} region indicate the phase space yielding $\tau_{\rm loss}<0.1~{\rm Gyr}$ {  via Parker wind mass-loss}.
In the bottom panel, we also plot the super-puff candidates listed in \citet{Chachan+20} with adding Kepler-90g \citep{Liang+20}, where the shortened planet name (e.g., Kepler-51b as K-51b) is denoted. 
}
\label{fig:Radius_ratio}
\end{figure}
%%%%%%%%%%%%%%%%%%%%%%%%%%%%%%%%%%%

First of all, we study a mass-radius relation of low-mass exoplanets to better understand how the atmospheric dust operates on the observable radius, motivated by \citet{Gao&Zhang20}.
We construct an interior structure model to calculate the mass-radius relation, as described in Appendix \ref{appendix:MR}.
Figure \ref{fig:MR_diagram} shows the planetary radius as a function of the planetary mass for different valuues of the atmospheric mass fraction $f_{\rm atm}$ and planetary intrinsic temperature $T_{\rm int}$.
Previous thermal evolution models predicted the intrinsic temperature of $\sim30~{\rm K}$ for old ($>3~{\rm Gyr}$) and $\sim70~{\rm K}$ for young ($0.1$--$3~{\rm Gyr}$) planets with $5M_{\rm \oplus}$, though it depends on the planetary mass and atmospheric mass fraction \citep{Lopez&Fortney14,Fortney+21}. 
Much higher values of $T_{\rm int}$ may be possible according to the retrieval study on warm Neptune GJ436b \citep{Morley+17}.
We test two atmospheric pressure of $P={10}^{-2}$ and ${10}^{-8}~{\rm bar}$ to define the observable radius. 
The former and the latter represent the radii of dust-free and highly dusty atmospheres, respectively.
In general, the radius increases with increasing the atmospheric mass fraction and the intrinsic temperature.
The radii rise toward smaller masses owing to weaker gravitational binding, while the radii are insensitive to $M_{\rm p}$ for higher mass planets.
These trends are in agreement with previous studies \citep[e.g.,][]{Rogers+11,Howe+14,Lopez&Fortney14,Gao&Zhang20}.
Importantly, the difference between the radii at ${10}^{-8}~{\rm bar}$ and ${10}^{-2}~{\rm bar}$ is small in the majority of thee mass space, especially at large planetary masses. 
This is because the pressure scale height is smaller for heavier planets.
Thus, atmospheric dust hardly influences the observable radii of hot/warm Jupiters, as already noticed by previous studies using interior structure models \citep{Baraffe+03,Burrows+03}.
At the small masses, the radius at $P={10}^{-8}~{\rm bar}$ is quite larger than that at $P={10}^{-2}~{\rm bar}$ owing to the large scale height.
However, the large scale height also increases the density at the sonic radius and thus the mass-loss rate via Parker wind.
The low-mass planets whose radii are sensitive to the pressure level tend to lose their atmospheres within a very short timescale, say $<0.1~{\rm Gyr}$.

From these arguments, we can arrive at one hypothesis: super-puffs are uncommon because atmospheric dust can drastically increase the observable radius only when a planet is near the stability limit.
Figure \ref{fig:Radius_ratio} shows the ratio of the ${10}^{-8}~{\rm bar}$ radius to the ${10}^{-2}~{\rm bar}$ radius, which diagnoses how much the atmospheric dust can enhance the radius, for various $f_{\rm atm}$, $T_{\rm int}$, and zero-albedo equilibrium temperature $T_{\rm eq}$.
As expected, the radius enhancement is remarkable only for a limited mass range.
To achieve the enhancement of $R({10}^{-8}~{\rm bar})/R({10}^{-2}~{\rm bar}){>}1.5$, for example, a planetary mass must be close to a threshold mass below which a catastrophic atmospheric escape takes place via Parker wind.
Note that our argument is insensitive to the choice of $\tau_{\rm loss}=0.1~{\rm Gyr}$ because the timescale of Parker wind mass-loss exponentially varies with a planetary mass.
For $T_{\rm int}=40~{\rm K}$ and $T_{\rm eq}=500~{\rm K}$ (top panel of Figure \ref{fig:Radius_ratio}), it is possible to achieve the radius enhancement more than $50\%$ only for planetary masses of ${\sim}2$--$4M_{\rm \oplus}$, depending on the atmospheric mass fraction.
This mass condition shifts to higher masses as the planetary intrinsic and equilibrium temperature rise, while the range of the $50\%$ radius enhancement remains narrow and always close the evaporation threshold (middle and bottom panels of Figure \ref{fig:Radius_ratio}).
When assuming $T_{\rm int}=75~{\rm K}$ and $f_{\rm atm}=0.05$, representative value for young ($<3$ Gyr) low-mass planets, more than $50\%$ radius enhancement can occur the mass range of $\sim2$--$3M_{\rm \oplus}$ for $T_{\rm eq}=300~{\rm K}$, $\sim2.5$--$4M_{\rm \oplus}$ for $T_{\rm eq}=500~{\rm K}$, and $\sim3.5$--$5M_{\rm \oplus}$ for $T_{\rm eq}=700~{\rm K}$.

An important implication of Figure \ref{fig:Radius_ratio} is that the radius enhancement is noticeable only near the stability limit even if we artificially impose the opaque pressure level at high altitudes.
\citet{Gao&Zhang20} attributed the significant radius enhancement for planets with atmospheric lifetimes of $0.1$--$1~{\rm Gyr}$ to enhanced haze opacity.
However, Figure \ref{fig:Radius_ratio} indicates that the enhanced haze opacity may not be the singular source of radius enhancement because the dust hardly enhances the radius of stable planets even if the opaque layer by dust is artificially imposed at high altitudes.

Therefore, atmospheric dust can render a planet a super-puff only when the planetary mass is close to the stability limit; otherwise, the dust hardly affects the planetary radius.
This can also be understood from the hydrostatic balance of an isothermal atmosphere that yields
\begin{eqnarray}\label{eq:main_ratio}
\nonumber
    \frac{d\ln{P}}{dr}&=&-\frac{GM_{\rm p}m_{\rm g}}{r^2k_{\rm B}T}\\
    \frac{R({10}^{-8}~{\rm bar})}{R({10}^{-1}~{\rm bar})}&=&\left[1-\frac{R({10}^{-1}~{\rm bar})k_{\rm B}T}{GM_{\rm p}m_{\rm g}}\ln{\left( \frac{{10}^{-1}~{\rm bar}}{{10}^{-8}~{\rm bar}}\right)}\right]^{-1}\\
    \nonumber
    &\approx&\left[1-\frac{7}{\Lambda}\left(\frac{2.35~{\rm amu}}{m_{\rm g}}\right)\right]^{-1},
%    \frac{d\ln{P}}{dr}&=&-\frac{GMm_{\rm g}}{r^2k_{\rm B}T}\\
%    \frac{R({10}^{-8}~{\rm bar})}{R({10}^{-2}~{\rm bar})}&=&1+\frac{R({10}^{-2}~{\rm bar})k_{\rm B}T}{GMm_{\rm g}}\ln{\left( \frac{{10}^{-2}~{\rm bar}}{{10}^{-8}~{\rm bar}}\right)}\\
%    \nonumber
%    &\approx&1+\frac{6}{\Lambda}\left(\frac{m_{\rm g}}{2.35~{\rm amu}}\right)^{-1},
\end{eqnarray}
where we have introduced the restricted Jeans parameter defined as \citep{Fossati+17}
\begin{equation}
    \Lambda \equiv \frac{GM_{\rm p}m_{\rm H}}{R({10}^{-1}~{\rm bar})k_{\rm B}T_{\rm eq}},
\end{equation}
where $m_{\rm H}$ is the mass of a hydrogen atom.
We note that Equation \eqref{eq:main_ratio} is not applicable to $\Lambda \la 7$ in which $10^{-8}~{\rm bar}$ exceeds the pressure at $r=\infty$ in the hydrostatic atmosphere assumed here.
Previous studies showed that planets with $\Lambda\la20$ undergo significant mass-loss via Parker wind and tend to lost their atmospheres \citep{Owen&Wu16,Fossati+17,Kubyshkina+18}.
The essence of Equation \eqref{eq:main_ratio} is that the degree of the radius enhancement is controlled by the ratio of pressure scale height to planetary radius, $rk_{\rm B}T/GMm_{\rm g}$.
Since atmospheric dust only operates on the logarithmic factor of Equation \eqref{eq:main_ratio}, large scale height to radius ratio is necessary for significant radius enhancement.
However, such large scale height inevitably leads to a high atmospheric density at a sonic point, resulting in significant atmospheric escape. 
Equation \eqref{eq:main_ratio} demonstrates that the elevated pressure level significantly affects the observed radius only for small $\Lambda$; for instance, ${\ga}50\%$ radius enhancement requires $\Lambda{\la}21$, which coincides with the boil-off threshold of $\Lambda\la20$.
%which is well below the boil-off threshold of $\Lambda\approx20$.
If we assume the atmospheric thickness is thinner than the core radius $R_{\rm core}$, i.e., $R_{\rm p}-R_{\rm core}{\la}R_{\rm core}$ that is expected as a consequence of spontaneous mass-loss after disk dispersal \citep{Ginzburg+16,Ginzburg+18}, the planetary mass range corresponding to $\Lambda\sim20$ can be crudely estimated from
\begin{equation}
    20\la\frac{GM_{\rm p}m_{\rm H}}{R_{\rm core}k_{\rm B}T_{\rm eq}}\la40.
\end{equation}
Using the rocky core radius relation of $R_{\rm core}=R_{\rm \oplus}(M_{\rm core}/M_{\rm \oplus})^{1/4}$ \citep{Lopez&Fortney14} and assuming $M_{\rm p}\approx M_{\rm core}$, the mass range responsible for the drastic radius enhancement might be evaluated as
\begin{equation}\label{eq:mass_range}
    2\la \left( \frac{M_{\rm p}}{M_{\rm \oplus}}\right) \left( \frac{T_{\rm eq}}{500~{\rm K}}\right)^{-4/3} \la5.
\end{equation}
Thus, we predict that planets of mass $\sim2$--$5M_{\rm \oplus}$ are potentially super-puffs owing to small $\Lambda$.
Despite the crudeness of the above estimation, Equation \eqref{eq:mass_range} reasonably explains the mass range responsible for the drastic radius enhancement in Figure \ref{fig:Radius_ratio}. 

To summarize, only planets on the verge of total atmospheric loss can benefit from the radius enhancement by high-altitude hazes.
This conclusion hardly depends on the choice of pressure level because of its logarithmic dependence. 
We suggest that this strict condition may be a reason why super-puffs are uncommon despite the fact that photochemical hazes are likely universally present.
Many super-puff candidates are indeed lie near the evaporation limit in $M_{\rm p}$-$T_{\rm eq}$ space (the bottom panel of Figure \ref{fig:Radius_ratio}, with assumptions of $f_{\rm atm}=0.05$ and $T_{\rm int}=75~{\rm K}$). 
We note that Figure \ref{fig:Radius_ratio} is merely representative, and the actual result depends on the atmospheric mass and intrinsic temperature that differ from planet to planet.
For example, Kepler-11f is placed in an unstable region in Figure \ref{fig:Radius_ratio}, but the planet presumably possesses an atmosphere given its old system age \citep[$8.5~{\rm Gyr}$,][]{Lissauer+13} and thus low intrinsic temperature. 

Lastly, we suggest that atmospheric dust is unlikely to be the direct cause of the large radii of super-puffs with $M_{\rm p}\ga10M_{\rm \oplus}$, such as Kepler-90g and WASP-107b.
The large radii of these massive low-density planets, or low-density sub-Saturns, are presumably caused by other reasons, such as a high atmospheric fraction \citep[as suggested for WASP-107b,][]{Piaulet+20}, high intrinsic temperature due to tidal heating \citep{Millholland19}, and circumplanetary ring \citep{Piro&Vissapragada19}.

\section{Discussion}\label{sec:discussion}
\subsection{Implications for Young Exoplanet Observations}

%We expect that more super-puffs can be found in young exoplanetary system. 
There may be more super-puffs in young stellar systems.
This is because planets have experienced only short age and have more chances to retain their primordial atmospheres.
%This is because planets have more chance to retain their atmospheres until the system age.
In addition, the bolometric energy-limited mass-loss implies the mass-loss timescale cannot be shorter than $\sim10~{\rm Myr}$ when the equilibrium temperature is low, say $T_{\rm eq}\la300~{\rm K}$, and/or the energy conversion efficiency is low (see Equation \ref{eq:bolo}).
This is comparable to the system ages of the youngest known exoplanet groups \citep{Mann+16b,David+19,Plavchan+20,Rizzuto+20}.
In such very young systems, relatively cool exoplanets might retain their atmospheres even if they seem unstable for a given $\Lambda$.
This further enhances the possibility that the radii of young exoplanets are inflated by atmospheric dust.
Some young super-puffs would eventually lose their atmospheres and become nominal super-Earths/sub-Neptunes.

The study of exoplanets in young stellar clusters and moving groups is a growing field.
Several young exoplanets with system ages of $\sim10~{\rm Myr}$ have already been discovered thus far, such as AU Mic b \citep{Plavchan+20,Hirano+20}, K2-33b \citep{Mann+16b}, V1298 Tau b \citep{David+19}, and HIP 67522b \citep{Rizzuto+20}. 
It is also worth noting that several studies reported that some young planets, namely K2-25b, K2-33b, and K2-95b, exhibit unusually large transit radii \citep{Mann+16,Mann+16b,Obermeier+16}.
The observed radii of these low-density young planets may be influenced by ongoing atmospheric escape entraining photochemical hazes and/or meteoric dust.
Transmission spectroscopy would be able to test this idea by investigating if the spectrum is featureless and exhibits a broad spectral slope (Section \ref{sec:spectrum}).
{Indeed, \citet{Thao+20} recently reported a featureless transmission spectrum for a young ($650$ Myr) Neptune-sized exoplanet, K2-25b, which may imply the presence of atmospheric dust.}
It is also expected that the contribution of meteoric dust becomes relatively high for young planetary systems owing to the remaining debris disks.
Observations by JWST-MIRI may be able to constrain how much meteoric dust is injected into the atmospheres of young exoplanets by searching for silicate feature at $\lambda\sim10~{\rm {\mu}m}$, as discussed in Section \ref{sec:spectrum}.

{ 
\subsection{Effects of Atmospheric Circulation}\label{sec:Kzz}
%%%%%%%%%%%%%%%%%%%%%%%%%%%%%%%%%%%
\begin{figure}[t]
\centering
\includegraphics[clip, width=\hsize]{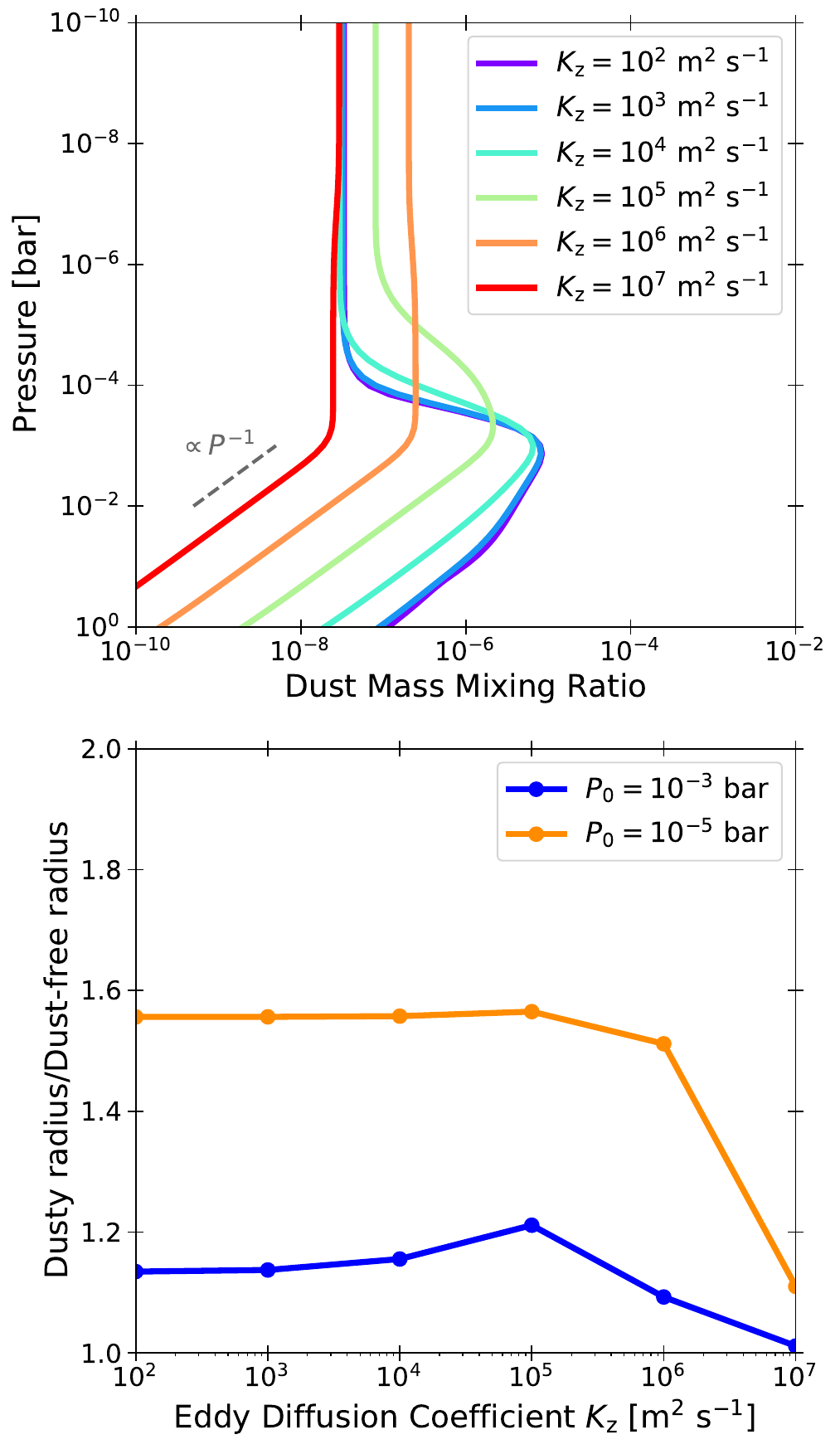}
\caption{  
(Top) Vertical distributions of dust mass mixing ratio. Different colored lines show the profile for different eddy diffusion coefficient $K_{\rm z}$. We set $\tau_{\rm loss}=1~{\rm Gyr}$, $P_{\rm 0}={10}^{-3}~{\rm bar}$, and $f_{\rm dust}={10}^{-4}$.
(Bottom) Effective transit radii of dusty atmospheres as a function of $K_{\rm z}$. The radii are normalized by the dust-free radius. The blue and orange lines show the radii for $P_{\rm 0}={10}^{-3}$ and ${10}^{-5}~{\rm bar}$, respectively.
We assume soot optical constants, $\tau_{\rm loss}=1~{\rm Gyr}$, and $f_{\rm dust}={10}^{-4}$.
}
\label{fig:Kzz}
\end{figure}
%%%%%%%%%%%%%%%%%%%%%%%%%%%%%%%%%%%

Thus far, we have focused on the ability of atmospheric escape to transport atmospheric dust to upper atmospheres.
In reality, however, there is an atmospheric circulation flow that is not escaping from the planet but still acts to transport dust.
Tracer transport by atmospheric circulation has been studied by global circulation models \citep{Parmentier+13,Charnay+15a,Zhang&Showman18a,Zhang&Showman18b,Komecek+19,Steinrueck+20}, although these studies focused on planets without atmospheric escape. 
For escaping atmosphere, \citet{Wang&Dai19} found the presence of circulating streamlines that eventually fall back to the planet.
Such a circulation does not contribute to the atmospheric loss yet may act to transport dust in vertical direction.
%to upper atmospheres.
%Thus, atmospheric circulation may play vital roles for transporting dust to upper atmospheres with keeping low mass-loss rate.

Here, we investigate how the dust distribution depends on the eddy diffusion coefficient $K_{\rm z}$, since the vertical transport by atmospheric circulation has been conventionally approximated by the eddy diffusion in a 1D framework \citep[e.g.,][]{Lavvas&Koskinen17,Kawashima&Ikoma18,Kawashima&Ikoma19,Ohno&Okuzumi18,Gao&Benneke18,Powell+18,Ormel&Min19,Ohno+20a}.
%The proper value of $K_{\rm z}$ has been uncertain.
For the equilibrium temperature relevant to super-puffs, say $T_{\rm eq}\sim500~{\rm K}$, \citet{Charnay+15a} and \citet{Komecek+19} derived the coefficient of $K_{\rm z}\sim{10}^3$--${10}^{5}~{\rm m^2~s^{-1}}$ using global circulation models with passive tracers, although it is unclear if the value is appropriate for super-puffs.
Thus, we examine the dust distributions for various values of $K_{\rm z}$.
Although the validity of the eddy diffusion approximation is still under debate \citep{Zhang&Showman18a,Zhang&Showman18b}, more detailed investigations with multidimensional hydrodynamical models including aerosol microphysics \citep{Lee+16,Lines+18} are beyond the scope of this study and are subjects for future studies.

%Moderate values of $K_{\rm z}$ help the upward transport of dust, while high values of $K_{\rm z}$ rather reduces the dust abundance in the outflow.
%Our conclusions would hold even if the vertical transport by atmospheric circulation is taken into account.
The top panel of Figure \ref{fig:Kzz} shows the vertical dust mass distributions for various $K_{\rm z}$.
We set $\tau_{\rm loss}=1~{\rm Gyr}$, $P_{\rm 0}={10}^{-3}~{\rm bar}$, and $f_{\rm dust}={10}^{-4}$. 
The vertical distribution is almost independent of the eddy diffusion for $K_{\rm z}\la{10}^{3}~{\rm {m}^2~s^{-1}}$.
The eddy diffusion starts to modify the vertical distributions at $K_{\rm z}\ga{10}^{4}~{\rm {m}^2~s^{-1}}$, especially in lower atmospheres. 
The mass mixing ratio at $P\ge P_{\rm 0}$ is then inversely proportional to $P$ and $K_{\rm z}$, as expected for diffusion-dominated downward transport with constant $K_{\rm z}$ \citep{Ohno&Kawashima20}.
At upper atmospheres with $P<P_{\rm 0}$, the mass mixing ratio initially increases with increasing $K_{\rm z}$ thanks to the diffusion transport.
However, as $K_{\rm z}$ further increases, the mixing ratio rather decreases with increasing $K_{\rm z}$.
This is because efficient eddy diffusion quickly removes the dust from the production region.
The bottom panel of Figure \ref{fig:Kzz} shows the resulting effective transit radius for various $K_{\rm z}$.
The effects of $K_{\rm z}$ on the observed radius is noticeable only for $K_{\rm z}\ga{10}^{5}~{\rm m^2~s^{-1}}$ and mostly acts to reduce the radius.
We note that \citet{Gao&Zhang20} obtained the same results for the $K_{\rm z}$ dependence (see their Section 4.2.1).
Thus, we predict that our conclusions hold even if the vertical transport by atmospheric circulation is taken into account.
%Thus, our conclusions would hold even if the vertical transport by atmospheric circulation is taken into account.
}

\subsection{Effects of Particle Porosity}\label{sec:porosity}
In this study, we have assumed that atmospheric dust particles are compact spheres.
However, recent studies have discussed that atmospheric dust particles are potentially porous aggregates---clusters of numerous tiny particles \citep{Marley+13,Ohno&Okuzumi18,Adams+19,Lavvas+19,Ohno+20a,Samra+20}.
To investigate possible impacts of this, we have performed a few simulations that take into account the porosity evolution of dust aggregates with a model of \citet{Ohno+20b}.
{In a nutshell, the porosity evolution lowers the density of dust particles through the growth.
The bulk volume of an aggregate can be expressed by
\begin{equation}
    V_{\rm agg}=\frac{mV_{\rm mon}}{m_{\rm mon}\phi},
\end{equation}
where $m_{\rm mon}$ and $V_{\rm mon}$ are the mass and volume of the smallest particles that constitute a dust aggregate---the so-called monomers.
The term $\phi$ is the volume filling factor defined as the ratio of the mean internal density of an aggregate to that of the individual monomer.
Conventionally, the shape of the aggregate is also characterized by the fractal dimension $D_{\rm f}$, defined as
\begin{equation}
    \frac{m}{m_{\rm mon}}=k_{\rm 0}\left( \frac{a_{\rm agg}}{a_{\rm mon}}\right)^{D_{\rm f}},
\end{equation}
where $k_{\rm 0}$ is the order unity prefactor, $a_{\rm mon}$ is the monomer's radius, and $a_{\rm agg}=(3V_{\rm agg}/4\pi)^{1/3}$ is the characteristic radius of a dust aggregate.
%$D_{\rm f}$ diagnoses the aggregate morphology; for example, spherical aggregates have $D_{\rm f}\approx3$, and plane-like aggregates have $D_{\rm f}\approx2$.
The growth via aggregate-monomer collisions leads to the formation of a sphere-like aggregate with $D_{\rm f}\approx3$, while the aggregate-aggregate collisions form a plane-line aggregate with $D_{\rm f}\approx2$ \citep[e.g.,][]{Meakin91,Cabane+93,Okuzumi+09}.
It has been known that photochemical hazes on Titan's upper atmosphere have $D_{\rm f}\approx 2$ \citep{Rannou+97}.

%%%%%%%%%%%%%%%%%%%%%%%%%%
\begin{figure}[t]
\centering
\includegraphics[clip, width=\hsize]{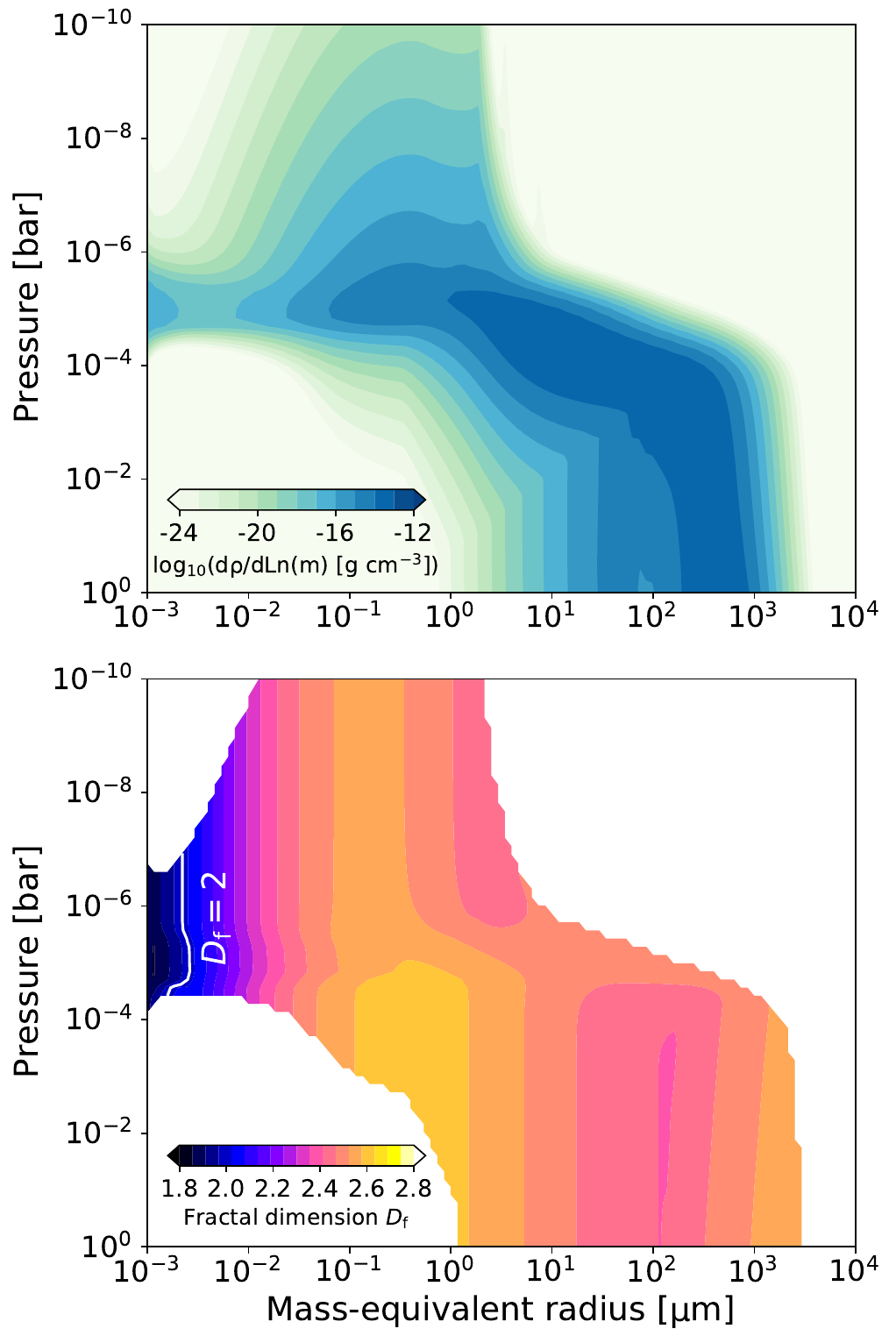}
\includegraphics[clip, width=\hsize]{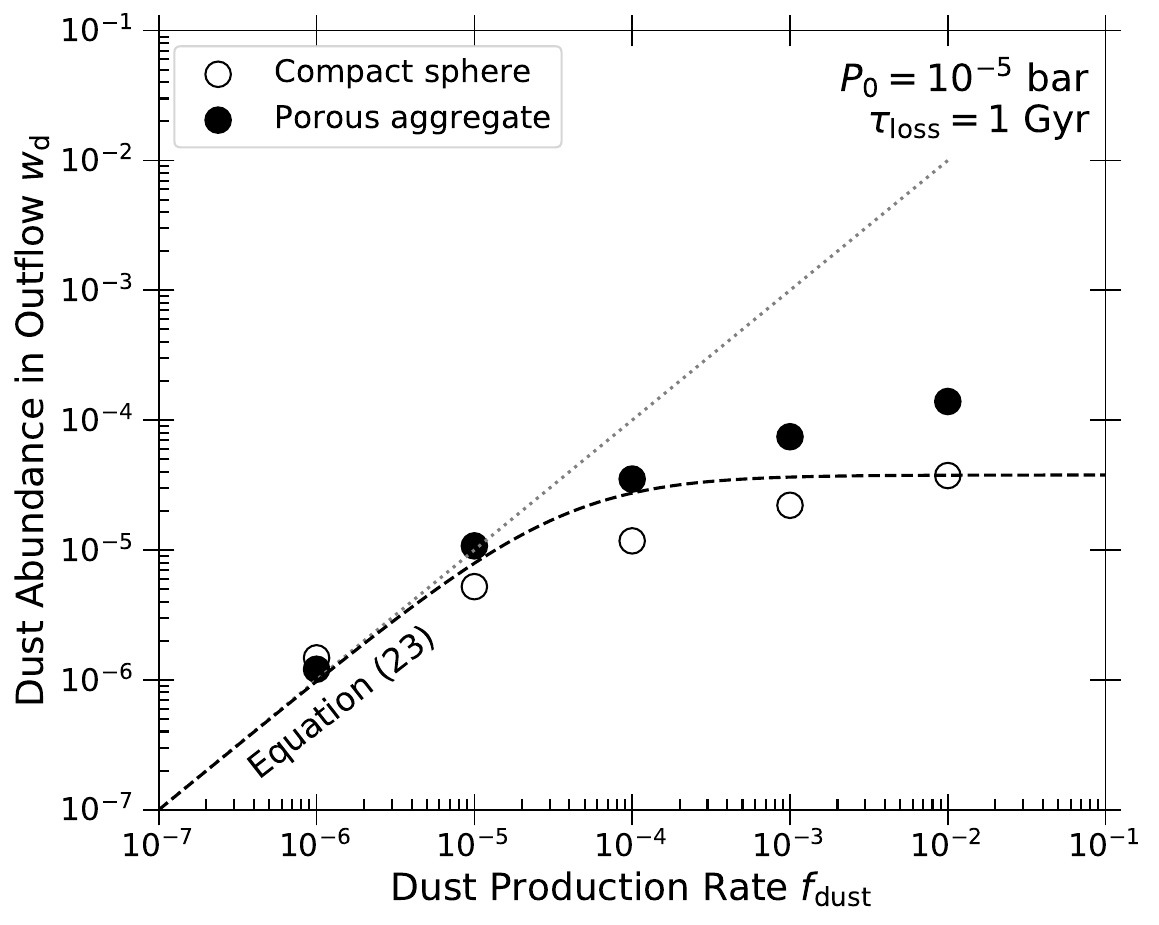}
\caption{(Top) Vertical mass distributions of porous aggregates as in Figure \ref{fig:dust_distribution}. The horizontal axis is the mass-equivalent radius, $a_{\rm 0}(m/m_{\rm 0})^{1/3}$, with $a_{\rm 0}=1~{\rm nm}$. We set $P_{\rm 0}={10}^{-5}~{\rm bar}$, $f_{\rm dust}={10}^{-3}$, and $\tau_{\rm loss}=1~{\rm Gyr}$. (Middle) Vertical distributions of particle fractal dimension $D_{\rm f}$ for $k_{\rm 0}=1$. (Bottom) Dust abundance in the outflow as in Figure \ref{fig:wd_summary}. The filled and empty dots indicate the results for aggregate and sphere models, respectively.
}
\label{fig:aggregate}
\end{figure}
%%%%%%%%%%%%%%%%%%%%%%%%%%%%%%%%%%%

One can apply the same formulation of a collision Kernel for spheres to aggregates using the characteristic radius $a_{\rm agg}$ \citep[e.g.,][]{Cabane+93}. 
The terminal velocity can be similarly computed from the characteristic radius, although a caution is needed to evaluate the geometric cross section of aggregates for $D_{\rm f}<2$ \citep{Okuzumi+09,Ohno+20b,Tazaki21}. 
It is worth noting that the terminal velocity is almost independent of the aggregate's characteristic size for $D_{\rm f}\approx 2$ in the free molecular flow regime, as the velocity is proportional to the particle mass-to-area ratio that is invariant with growth for $D_{\rm f}\le2$ \citep[see, e.g., Equation (26) of][]{Ohno+20b}.

Previous studies usually assumed a fixed value of $D_{\rm f}=2$ through the simulations \citep[e.g.,][]{Gao+17,Lavvas+19}; however, this approach is not adequate for dust aggregates in escaping atmospheres.
This is because, as the terminal velocity is invariant with aggregate sizes for $D_{\rm f}=2$, the dust aggregates are always transported upward by outflow even if they grow into extremely large sizes, unless the size becomes larger than the mean free path of ambient gas particles.
In reality, large aggregates tend to experience collisions with smaller aggregates, leading to increase $D_{\rm f}$ and terminal velocity.
The model of \citet{Ohno+20b} adopts the volume averaging method of \citet{Okuzumi+09} to explicitly simulate the evolution of $V_{\rm agg}$ at each mass bin without assuming a specific $D_{\rm f}$.
This approach enables us to trace the evolution of $D_{\rm f}$ caused by collisions with small particles, which enables us to avoid the aforementioned unrealistic behaviors.
%This approach has an advantage that can trace which aggregate-aggregate collisions or aggregate-monomer collisions is dominant in a self-consistent manner.
}
%The model simulates the evolution of particle mean density in each mass bin following the volume averaging method of \citet{Okuzumi+09}.
We also include the compression of dust aggregates caused by ram pressure from gas drag by setting a minimum particle density following the equilibrium porosity recipe of \citet{Ohno+20a}.
For more detailed descriptions of the porosity evolution model, we refer readers to \citet{Ohno+20a,Ohno+20b} and references therein.

The porosity evolution hardly alters the qualitative results of this study.
Figure \ref{fig:aggregate} shows vertical size distributions (top), distributions of particle fractal dimensions (middle), and dust abundances at ${10}^{-9}~{\rm bar}$ (bottom), where we have included the eddy diffusion with $K_{\rm z}={10}^{4}~{\rm m^2~s^{-1}}$ to facilitate the convergence of deep atmosphere profiles.
Although outflow can transport dust aggregates with sizes much larger than compact dust, the dust eventually grows into too large size to be blown up by outflow (top panel).
This stems from the timescale of settling-driven collision growth that is independent of the internal density of dust particles (see Equation \ref{eq:tau_coal}).
As a result, the abundance of dust aggregates in escaping atmospheres is regulated by the growth and settling, similar to the results for compact dust (bottom panel).
While the porosity evolution does enhance the dust abundance at upper atmospheres by a factor of up to $\sim3$ from the compact sphere cases, the abundance can still be roughly explained by our analytical theory (Equation \ref{eq:wd_main}).

One of the interesting results is that dust aggregates have large sizes and moderately compressed internal structures ($D_{\rm f}{\sim}2.5$) in escaping atmospheres.
In particular, the obtained fractal dimension is quite larger than $D_{\rm f}=2$, which was usually assumed for uncompressed aggregates \citep[e.g.,][]{Gao+17,Lavvas+19,Ohno+20a}.
This is caused by efficient collisions between newly produced dust and growing aggregates around the dust production region.
Because outflow inhibits the settling of a dust aggregate until the settling velocity exceeds the outflow velocity, the dust experiences many collisions with newly produced tiny dust, leading to an increase in $D_{\rm f}$ by filling pores within an aggregate.
Once the settling becomes dominant, the fractal dimension starts to decrease via collisions between similar-sized aggregates, as tiny dust hardly enters lower altitudes due to the outflow.
Since the compressed aggregates can efficiently flatten the transmission spectra \citep{Adams+19}, the result potentially suggests that planets with atmospheric escape tend to exhibit the flat spectra.
We defer detailed investigations of the grain growth with porosity evolution and its observational implications to our future studies.

\subsection{Distinguishing from Circumplanetary Ring}\label{sec:ring}
There is an alternative hypothesis for the low-density nature of super-puff: the presence of circumplanetary rings \citep{Piro&Vissapragada19,Akinsanmi+20}.
The ring can cause additional occultation and overestimation of the transit radius, leading to low planetary density \citep{Zuluaga+15}.
Previous studies proposed that the careful analysis of photometric and/or spectroscopic light curve can detect exoplanetary ring \citep[e.g.,][]{Barnes&Fortney04,Ohta+09,Aizawa+17}.
{  For example, it is expected that ringed planets exhibit asymmetric light curves with a long duration for ingress and/or egress of planet's transit \citep{Aizawa+17}.
If the ring particles have a size parameter ($\sim a/\lambda$) comparable to the viewing angle ($R_{\rm s}/a_{\rm orb}$), the ring particles cause forward scattering that increases the photometric flux right before and after the planet's transit \citep{Barnes&Fortney04}.
}
The ring scenario also demands that the planet has a large obliquity---the angle between the planetary rotation axis and orbital axis---to produce a large transit depth \citep{Piro&Vissapragada19}. 
Thus, it may be possible to test the scenario by searching for obliquity signatures, which potentially appear in eclipse mapping \citep{Rauscher17} and thermal phase curve observations \citep{Ohno&Zhang19,Adams+19_obl}.

Here, we also suggest that the transmission spectrum will be useful to distinguish whether atmospheric dust or a circumplanetary ring causes the large radius of a super-puff.
To demonstrate it, we construct a simple model of the transmission spectra applicable to ringed planets.
We assume a horizontally uniform ring whose outer edge may be estimated as the Roche radius, given by \citep[e.g.,][]{Piro&Vissapragada19}
\begin{equation}
    R_{\rm out}\approx 2.46\left( \frac{3M_{\rm p}}{4\pi \rho_{\rm r}}\right)^{1/3}
\end{equation}
where $\rho_{\rm r}$ is the density of a ring particle.
Motivated by the fact that Saturnian ring particles are highly porous \citep[porosities of $50$--$90$\%,][]{Zhang+17}, we assume $\rho_{\rm r}=0.3~{\rm g~{cm}^{-3}}$ that corresponds to a silicate particle with $90$\% porosity.
This choice can satisfy the particle density required for most super-puffs in \citet{Piro&Vissapragada19}.
The inner ring edge is highly uncertain, and hence we simply assume $R_{\rm in}\approx 0.5R_{\rm out}$ following \citet{Piro18}, which is motivated by Saturn's ring.
For the sake of simplicity, we only assume a face-on ring whose rotation axis coincides with the line of sight.
This is obviously an idealized assumption since the axis is likely oblique to the line of sight in reality, but it does not alter our qualitative result discussed below.
In this context, from the modification of Equation \eqref{eq:R_eff}, the transmission spectrum of a ringed planet is given by
\begin{equation}
    R_{\rm eff}^2=R_{\rm 0}^2+2\int_{\rm R_{\rm 0}}^{R_{\rm H}}[1-\exp(-\tau_{\rm s})\exp{(-\tau_{\rm r}\mathcal{H}((r-R_{\rm in})(R_{\rm out}-r))}]rdr,
\end{equation}
where $\tau_{\rm r}$ is the nadir optical depth of the ring, and $\mathcal{H}(x)$ is the Heaviside step function defined as $1$ at $x>0$ and $0$ at $x<0$.
We assume a gray ring opacity, as ring particles smaller than ${\sim}100\ {\rm cm}$ would be removed by the orbital decay through the Pointing-Robertson drag \citep{Schlichting&Chang11}.

%%%%%%%%%%%%%%%%%%%%%%%%%%
\begin{figure}[t]
\centering
\includegraphics[clip, width=\hsize]{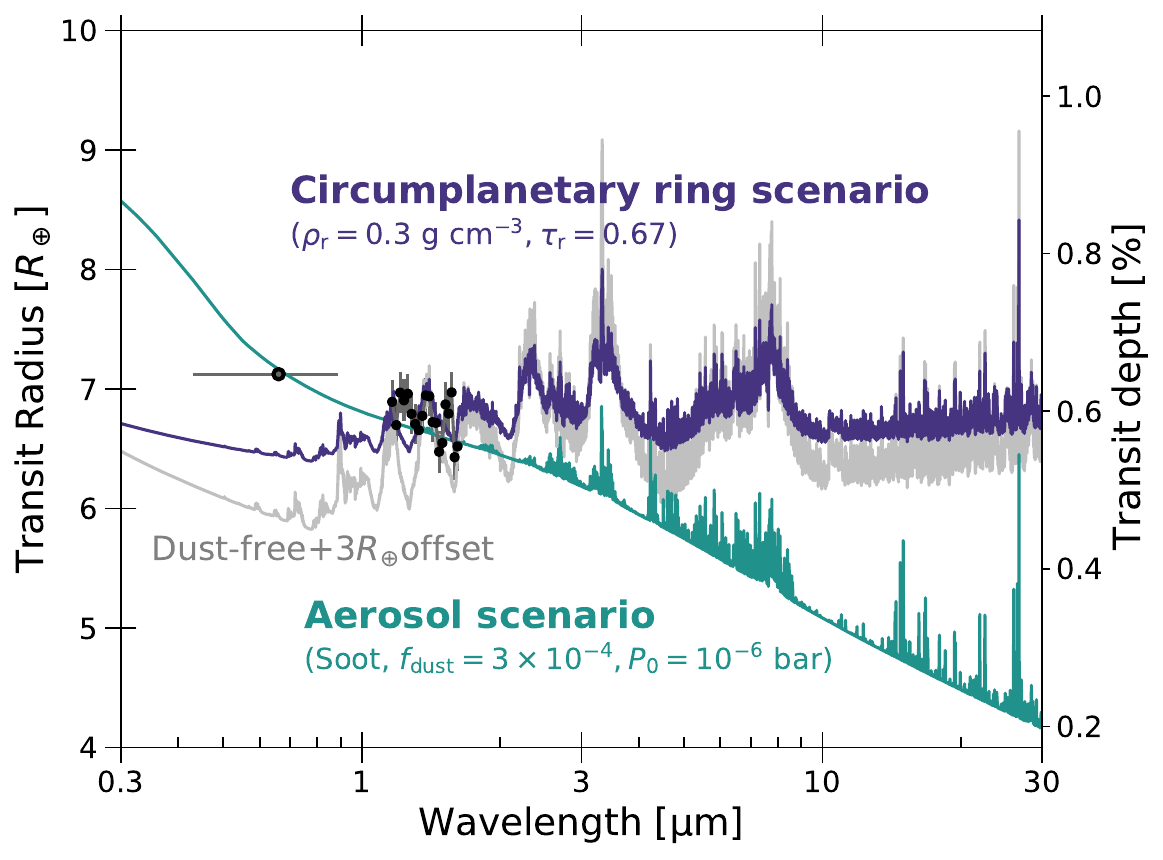}
\caption{An anticipated difference of transmission spectra between super-puffs inflated by dusty atmospheres (the green line, aerosol scenario) and circumplanetary rings (the purple line, ring scenario). For the ringed planet, we assume a face-on ring composed of porous silicate particles with internal density of $\rho_{\rm r}=0.3\ {\rm g\ {cm}^{-3}}$ and nadir optical depth of $\tau_{\rm r}=0.67$, a dust-free atmosphere, and $\tau_{\rm loss}=1~{\rm Gyr}$.
The gray line shows the dust-free and ring-free spectrum with $3R_{\rm \oplus}$ offset.
%Also shown are the radius of Kepler-51b observed by HST-WFC3 \citep{Libby-Roberts+20}.
}
\label{fig:ring}
\end{figure}
%%%%%%%%%%%%%%%%%%%%%%%%%%%%%%%%%%%

%We predict that ringed super-puffs exhibit transmission spectra significantly different from those of super-puffs with dusty atmospheres.
The shape of the transmission spectrum of a ringed super-puff is significantly different from that of a super-puff with a dusty atmosphere. 
Figure \ref{fig:ring} shows the spectrum of a ringed planet with $M_{\rm p}=3.5M_{\rm \oplus}$, where $\tau_{\rm r}$ is adjusted to match the near-infrared radius of Kepler-51b.
The ringed planet exhibits a spectrum with reduced spectral features as compared to that of a ring-free planet (see the gray line in Figure \ref{fig:ring}).
The spectral features still appear because the atmosphere inside the ring's inner edge is still visible to an observer.
The features would be further diminished if the total atmospheric mass is low, the inner ring edge is closer to the planet, or the ring has an inclination to block the transmitted starlight from the atmosphere.
Importantly, the observed radii of ringed super-puffs would not monotonically decrease with increasing wavelength because ring particles should be much larger than atmospheric dust.
Small ring particles might exist if the ring is optically thick \citep{Schlichting&Chang11}. 
In that case, the ring would produce silicate features that can be used to diagnose the presence of the ring.
The two scenarios yield a difference of $\sim1000~{\rm ppm}$ in the transit depth of a Kepler-51b like planet, which would be distinguished by the future observations of ground-based telescopes as well as by JWST.
In the specific case of Kepler-51b, existing visible and near-infrared observations are better explained by the atmospheric dust scenario, although a uniform data analysis may be needed to avoid the possible offsets between different observations.

%%%%%%%%%%%%%%%%%%%%%%%%%%%%
{ 
\subsection{Dust Fragmentation}\label{sec:fragment}
\begin{figure}[t]
\centering
\includegraphics[clip, width=\hsize]{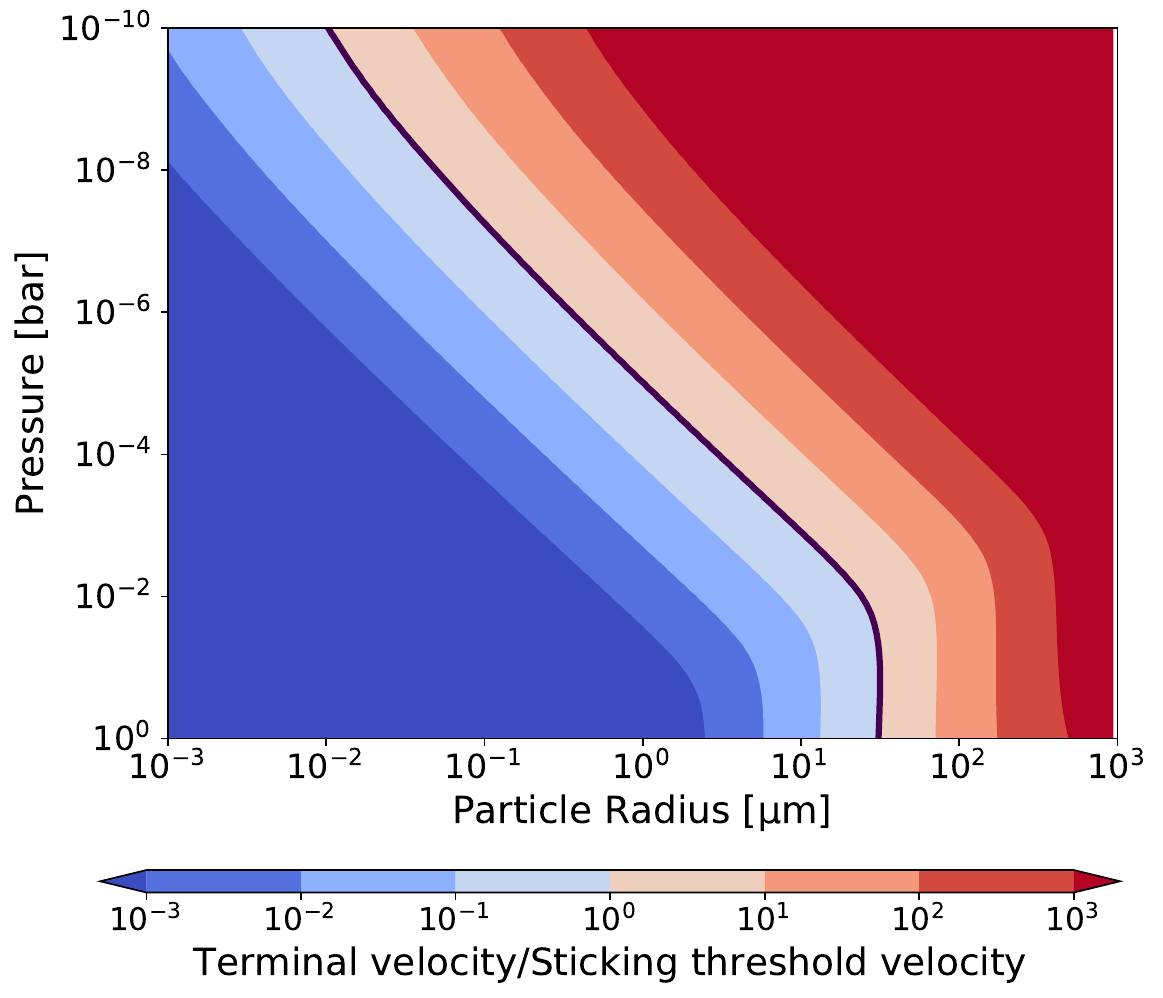}
\caption{  
The ratio of particle terminal velocity to sticking threshold velocity (Equation \ref{eq:v_cri}), $v_{\rm d}/v_{\rm cri}$, for an equal-sized collision as a function of pressure and particle radius.
The black line denotes the threshold radius of $v_{\rm d}=v_{\rm cri}$ below which the collision sticking can occur.
We assume $\gamma_{\rm surf}=70~{\rm mJ~{m}^{-2}}$, $E=10~{\rm GPa}$,  $\nu=0.3$, and $\tau_{\rm loss}=1~{\rm Gyr}$.
}
\label{fig:v_frag}
\end{figure}
%%%%%%%%%%%%%%%%%%%%%%%%%%%%%%%%%%%

We assumed perfect sticking for dust collisions; however, the assumption is invalid if the collision velocity is too fast.
In short, collision sticking can occur if the collision kinetic energy is smaller than the adhesion binding energy for two particles in contact. 
Based on the contact theory of elastic spheres \citep[the so-called the JKR theory,][]{Johnson+71}, the threshold collision velocity, below which the sticking can occur between particles 1 and 2, is given by \citep{Chokshi+93}
\begin{equation}\label{eq:v_cri}
    v_{\rm cri}\approx 3.86\frac{\gamma_{\rm surf}^{5/6}}{\mathcal{E}^{1/3}\mathcal{R}^{5/6}\rho_{\rm d}^{1/2}},
\end{equation}
where $\gamma_{\rm surf}$ is the surface energy, $\mathcal{R}=a_{\rm 1}a_{\rm 2}/(a_{\rm 1}+a_{\rm 2})$ is the reduced particle radius, and $\mathcal{E}=E/2(1-\nu^2)$ with Young's modulus $E$ and Poisson's ratio $\nu$.
We assume $\gamma_{\rm surf}=70~{\rm mJ~{m}^{-2}}$, $E=10~{\rm GPa}$, and $\nu=0.3$ in this section.
These values correspond to the material properties of graphite in the Table 2 of \citet{Chokshi+93}, and $\gamma_{\rm surf}$ and $\nu$ are comparable to those of Titan's haze analog produced in experiments \citep{Yu+17,Yu+20}.
We note that the haze analog has the Young's modulus lower than that assumed here \citep[$E=3~{\rm GPa}$,][]{Yu+17}.
The threshold velocity decreases with increasing the particle sizes because the surface to volume ratio decreases.
Higher surface energy and lower Young's modulus also facilitate the collision sticking by enhancing the adhesion force.

Since the thermal particle velocity is much slower than the sticking threshold velocity \citep{Lavvas&Koskinen17}, here we examine if particles can stick via collisions driven by differential gravitational settling. 
The collision velocity is comparable to the terminal velocity in this case, and thus the collision sticking can occur for $v_{\rm d}<v_{\rm cri}$.
We show the ratio of the terminal velocity to the threshold velocity $v_{\rm d}/v_{\rm cri}$ in Figure \ref{fig:v_frag}.
%The terminal velocity is independent of atmospheric pressure in the deep dense regions for which the Stokes's law can be applied.
The terminal velocity is always slower than the sticking threshold for $a\la30~{\rm {\mu}m}$ at $P\ga{10}^{-2}~{\rm bar}$.
The terminal velocity becomes comparable to the sticking threshold at smaller particle radii in lower pressure regions where the free molecular flow regime applies.
In particular, for $P\la{10}^{-8}~{\rm bar}$, the terminal velocity exceeds the sticking threshold at $a\la0.1~{\rm {\mu}m}$, which is smaller than the threshold size $a_{\rm cri}$ above which dust particles cannot be transported by outflow for $\tau_{\rm loss}=1~{\rm Gyr}$ (see Equation \ref{eq:a_cri}).
Thus, dust formed in such extreme upper atmospheres can be easily transported upward without growth thanks to inefficient collision sticking.
On the other hand, at deeper atmospheres, on which we have focused as dust production regions, the terminal velocity exceeds $v_{\rm cri}$ only at the size much larger than $a_{\rm cri}$, unless the mass-loss timescale is extremely short.
Since the dust particles quickly settle down to deeper regions once their sizes exceed $a_{\rm cri}$ (Figure \ref{fig:dust_distribution}), they hardly grow into the sizes for which the terminal velocity exceeds the sticking threshold.
Thus, the assumption of perfect sticking is mostly valid in the parameter spaces examined in this study.

\subsection{Can Dust Survive in Ultra-hot Outflow?}\label{sec:dust_survive}
We assumed that the escaping outflow has the planetary equilibrium temperature.
%However, the outflow temperature could differ from the equilibrium temperature in reality.
However, the outflow can be much hotter than the equilibrium temperature in certain cases; for example, XUV illumination \citep[e.g.,][]{Wang&Dai18,Wang&Dai19} as well as the dissipation of magnetohydrodynamic waves \citep{Tanaka+14,Tanaka+15} can heat upper atmospheres to temperature as high as ${10}^{4}~{\rm K}$.
This poses a natural question: can dust survive in the ultra-hot outflow without evaporation?

We first investigate whether the collision heating is fast enough to evaporate the dust.
The collision heating rate is given by \citep[Chapter 24 of][]{Draine11}
\begin{equation}
    \frac{dE_{\rm dust}}{dt}= \pi a^2\sqrt{\frac{8k_{\rm B}T_{\rm gas}}{\pi m_{\rm g}}} n_{\rm g}k_{\rm B}(T_{\rm gas}-T_{\rm dust}),
\end{equation}
where $E_{\rm dust}$ is the internal energy of dust, $n_{\rm g}$ is the number density of gas particles, and $T_{\rm dust}$ and $T_{\rm gas}$ are the dust and gas temperatures.
The timescale for heating up the dust to complete evaporation can be estimated by
\begin{equation}
    %\tau_{\rm heat}=\sqrt{\frac{\pi}{8}}\frac{4\rho_{\rm p}c_{\rm v}a}{3 n_{\rm g} c_{\rm s}k_{\rm B}},
    \tau_{\rm heat}=mL_{\rm v}\left( \frac{dE_{\rm dust}}{dt}\right)^{-1}\approx \sqrt{\frac{\pi }{8}}\frac{4a\rho_{\rm d}L_{\rm v}}{3 Pc_{\rm s}},
\end{equation}
where $L_{\rm v}$ is the latent heat of dust evaporation, and we assume $T_{\rm gas}\gg T_{\rm dust}$.
% and use the relation of $P=n_{\rm g}k_{\rm B}T_{\rm gas}$.
The latent heat is typically $L_{\rm v}\sim{10}^6$--${10}^{7}~{\rm J~{kg}^{-1}}$, depending on the substance \citep[e.g.,][]{Gao+20}.
The collision heating is fast enough if the heating timescale is faster than the transport timescale $\tau_{\rm tran}=H/v_{\rm g}$.
The ratio of the two timescales is given by
\begin{eqnarray}
%\nonumber
    \frac{\tau_{\rm heat}}{\tau_{\rm tran}}&=&\sqrt{\frac{\pi }{8}}\frac{4a\rho_{\rm d}L_{\rm v}v_{\rm g}}{3 HPc_{\rm s}}=\frac{a \rho_{\rm d}g^2L_{\rm v}}{\sqrt{72\pi}GP^2c_{\rm s} \tau_{\rm loss} }\\
    \nonumber
    &\approx& 0.3 \left( \frac{a}{ 0.1~{\rm {\mu}m}}\right)\left( \frac{P}{ {10}^{-8}~{\rm bar}}\right)^{-2} \left( \frac{g}{ {10}~{\rm m~s^{-2}}}\right)^{2}\left( \frac{\tau_{\rm loss}}{ 1~{\rm Gyr}}\right)^{-1}\\
    \nonumber
    &&\times \left( \frac{\rho_{\rm d}}{ 1~{\rm g~{cm}^{-3}}}\right)\left( \frac{L_{\rm v}}{ {10}^{7}~{\rm J~{kg}^{-1}}}\right)\left( \frac{c_{\rm s}}{ {10}^{4}~{\rm m~s^{-1}}}\right)^{-1},
\end{eqnarray}
where we have used $\dot{M}=4\pi r^2 \rho_{\rm g}v_{\rm g}$ for the second equation.
The strong heating of atmospheric gases occurs at $\la {10}^{-6}~{\rm bar}$ \citep[e.g.,][]{Koskinen+13}, while the heating is fast enough at $\ga{10}^{-8}~{\rm bar}$.
Thus, the collision heating potentially does matter, especially for tiny dust particles that can be transported by outflow.
%Solving $P$ at $\tau_{\rm tran}=\tau_{\rm eva}$ with using the relation of $\dot{M}=4\pi r^2 \rho_{\rm g}v_{\rm g}$, we obtain the threshold pressure level above which the dust can survive to the collision heating, given by

The above argument does not mean that dust cannot survive in the ultra-hot outflow, since the cooling process was neglected.
%In reality, the dust temperature is determined by the balance among radiation cooling, illumination heating, and collision heating.
%When the outflow is optically thin, the equilibrium dust temperature can be estimated as
If the collision heating dominates over the illumination heating, taking the balance between radiation cooling and collision heating, the equilibrium dust temperature can be estimated as
\begin{eqnarray}\label{eq:T_dust}
    \nonumber
    %T_{\rm dust}&=&\epsilon_{\rm s}^{-1/4} T_{\rm eq}\left( 1+\sqrt{\frac{k_{\rm B}T}{2\pi m_{\rm g}}} n_{\rm g}k_{\rm B}\frac{(T_{\rm gas}-T_{\rm dust})}{\epsilon_{\rm s}\sigma T_{\rm eq}^4}\right)^{1/4}\\
    %&\approx& \epsilon_{\rm s}^{-1/4}T_{\rm eq}\left( 1+\frac{c_{\rm s}P}{\sqrt{2\pi}\epsilon_{\rm s}\sigma T_{\rm eq}^4}\right)^{1/4},
    4\pi a^2 \epsilon_{\rm s} \sigma T_{\rm dust}^4&=& \pi a^2\sqrt{\frac{8k_{\rm B}T}{\pi m_{\rm g}}} n_{\rm g}k_{\rm B}(T_{\rm gas}-T_{\rm dust})\\
    T_{\rm dust}&\approx& \left( \frac{c_{\rm s}P}{\sqrt{2\pi}\epsilon_{\rm s}\sigma }\right)^{1/4},
    %\nonumber
    %&\approx& 290\epsilon_{\rm s}^{-1/4}~{\rm K} \left( \frac{c_{\rm s}}{{10}^4~{\rm m~s^{-1}}}\right)^{1/4}\left( \frac{P}{{10}^{-6}~{\rm bar}}\right)^{1/4}.
\end{eqnarray}
where $\epsilon_{\rm s}$ is the emissivity, defined by \citep{Draine11}
\begin{equation}
    \epsilon_{\rm s}(a,T)\equiv \frac{\int d\nu B_{\rm \nu}(T)Q_{\rm abs}(\nu,a)}{\int d\nu B_{\rm \nu}(T)},
\end{equation}
where $B_{\rm \nu}$ is the Plank function, and $Q_{\rm abs}$ is the absorption coefficient given by the Mie theory.
In principle, the more absorptive the particle is, the more efficiently the radiation cooling works.
%We note that the equilibrium dust temperature does not explicitly depend on the gas temperature, which is absorbed in atmospheric pressure.

%%%%%%%%%%%%%%%%%%%%%%%%%%%%%%%%%%%
\begin{figure}[t]
\centering
\includegraphics[clip, width=\hsize]{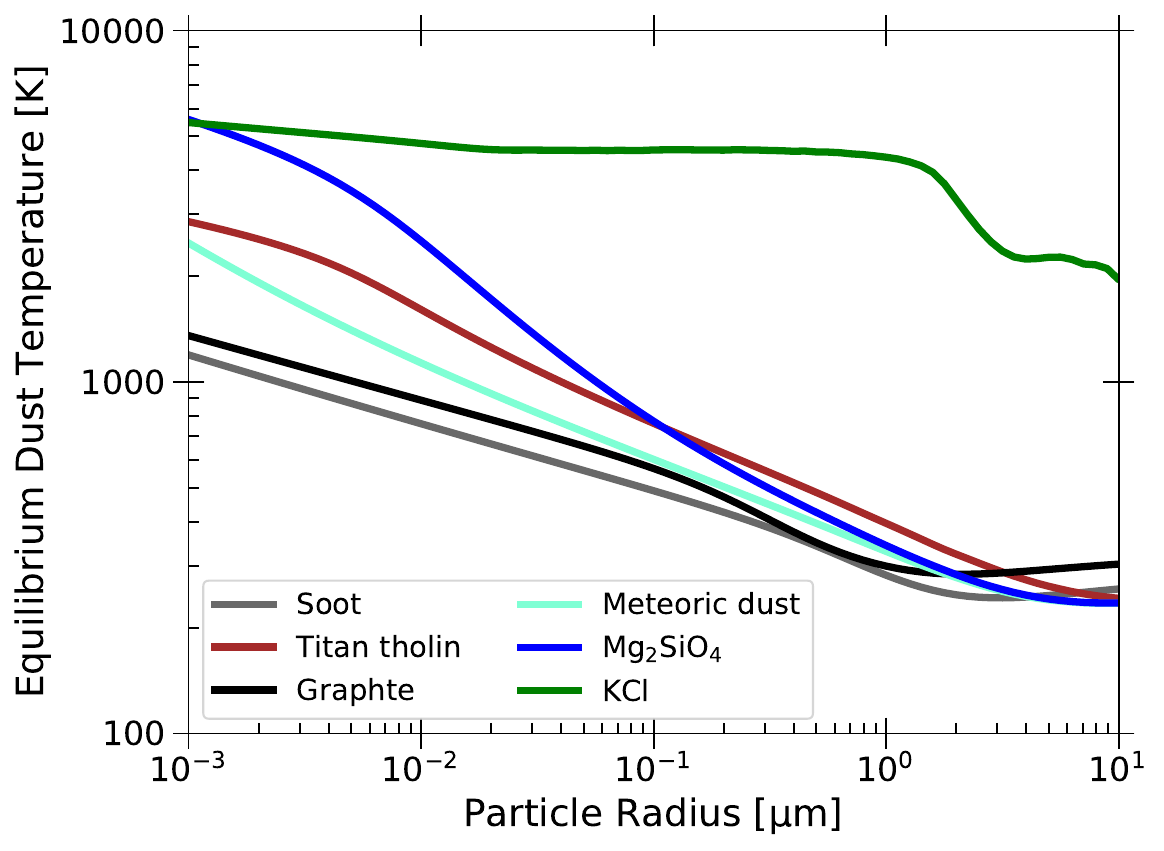}
\caption{  
Equilibrium dust temperature controlled by radiation cooling and heating by gas particle collisions. The horizontal axis is the particle radius.
Different colored lines show the equilibrium temperature for different optical constants, as presented in Figure \ref{fig:Mie_opacity}. We set $P={10}^{-6}~{\rm bar}$ and $T_{\rm gas}={10}^{4}~{\rm K}$.
}
\label{fig:T_dust}
\end{figure}
%%%%%%%%%%%%%%%%%%%%%%%%%%%%%%%%%%%

Whether the dust can survive or not depends on the emissivity, which turns out to depend on dust particle sizes and optical properties.
Figure \ref{fig:T_dust} shows the equilibrium dust temperature for several exoplanetary aerosol analogs.
%While large particles can retain cool temperature thanks to efficient radiation cooling, 
The collision heating significantly increases the temperature of small dust.
For example, for optical constants of Titan tholin, meteoric dust, and Mg$_2$SiO$_4$, the particles are heated to the temperature of $\gg1000~{\rm K}$ when their sizes are smaller than $\sim0.03~{\rm {\mu}m}$.
KCl condensates are always heated to an extremely high temperature because KCl is almost purely scattering material and hardly emits the deposited heat.
In these cases, dust is unlikely to survive in the upper atmospheres when the outflow is extremely hot.
On the other hand, dust made of absorbing materials, such as soot and graphite, likely survive in the ultra-hot outflow, as they can retain a temperature of $\la1000~{\rm K}$ even at $a\la0.01~{\rm {\mu}m}$ thanks to efficient radiation cooling.
% for more absorbing materials, such as soot and graphite.
%dust could retain the temperature of $\la1000~{\rm K}$ at smaller particle sizes for more absorbing materials, such as soot and graphite.
%Thus, we predict that dust can survive even in the ultra-hot outflow if they are made of absorbing materials.

%Thus, if the dusty outflow is indeed present, the dust would be made of absorbing materials.
%is uncertain for exoplanetary aerosols.
%For graphite-like compositions assumed in \citet{Wang&Dai19}, the emissivity at $\sim1000~{\rm K}$ is $\epsilon_{\rm s}\sim 0.3~(a/1~{\rm {\mu}m})$ according to Figure 24.3 of \citet{Draine11}.
%Assuming this relation, the equilibrium dust temperature may be estimated as
%\begin{eqnarray}\label{eq:T_dust2}
%    T_{\rm dust}&\approx& 1200~{\rm K} \left( \frac{c_{\rm s}}{{10}^4~{\rm m~s^{-1}}}\right)^{1/4}\left( \frac{P}{{10}^{-6}~{\rm bar}}\right)^{1/4}\left( \frac{a}{0.01~{\rm {\mu}m}}\right)^{-1/4}.
%\end{eqnarray}
%Photochemical haze formation was experimentally observed at the hot environment of $1473~{\rm K}$ \citep{Fleury+19}.
%Thus, we predict that the dust can survive in the hot outflow as long as $a\ga0.01~{\rm {\mu}m}$.

}
%%%%%%%%%%%%%%%%%%%%%%%%%%%%%%%%%
\section{Summary}\label{sec:summary}
In this study, we have studied the effects of a dusty outflow on the observable radius of low-mass exoplanets using a microphysical model of grain growth. 
We have examined a wide range of dust production rates and formation altitudes to discuss what kinds of aerosols are responsible for enhancing the observable radius.
Our findings are summarized as follows.

\begin{enumerate}

\item Collision growth of dust is usually efficient in escaping atmospheres, especially for high dust production rate (Section \ref{sec:result1}).
When the dust production rate is low, dust can be completely blown up to upper atmospheres, though it only yields a low dust abundance in escaping atmospheres.
However, high dust abundance causes efficient collision growth, leading to subsequent gravitational settling of dust.
For example, even if the dust production rate is so high that the mass mixing ratio reaches ${10}^{-2}$, the growth followed by the settling tends to reduce the mixing ratio by several orders of magnitude.
Thus, while a high dust abundance is favored to cause the inflation of observable radii, there is a dilemma that the high abundance enhances particle growth and hinders upward transport.

\item The dust abundance of the outflow also highly depends on the pressure level of the dust production region (Section \ref{sec:result_abundance}).
Intense atmospheric escape with short $\tau_{\rm loss}$ can facilitate the blow up of dust to upper atmospheres.
For example, in the case of $\tau_{\rm loss}=0.1~{\rm Gyr}$ and $f_{\rm dust}={10}^{-2}$, the dust abundance in the outflow reaches $3\times{10}^{-3}$ for $P_{\rm 0}={10}^{-6}~{\rm bar}$ while the abundance is ${10}^{-6}$ for $P_{\rm 0}={10}^{-2}~{\rm bar}$.

\item We have constructed a simple analytical model that predicts the dust abundance in escaping atmospheres (Equation \ref{eq:wd_main}, Appendix \ref{sec:anal_derivation0}).
The model can be easily utilized in hydrodynamical models as well as thermal evolution models of exoplanets with atmospheric escape.
In the limit of a high dust production rate, the model predicts that the dust abundance is regulated to be about the Mach number of the dust production region, which reasonably explains the numerical results.

\item The dusty outflow can cause a significant radius enhancement only when the dust is formed at high altitudes (Section \ref{sec:impact_radius}).
For example, the dusty outflow can cause the radius enhancement as large as a factor of $2$ or even more for $P_{\rm 0}={10}^{-6}~{\rm bar}$, whereas the outflow only enhances the radius by $\la10\%$ for any combination of $\tau_{\rm loss}$ and $f_{\rm dust}$ when $P_{\rm 0}={10}^{-2}~{\rm bar}$.

\item Since the high altitude of dust production region is required, we have suggested that photochemical haze is a promising candidate to generate the dusty outflow that may explain the large radii of super-puffs (Section \ref{sec:dis_dust_nature}).
This supports the idea of \citet{Gao&Zhang20}.
Alternatively, meteoric dust might also work to enhance the observable radii if a sufficient amount of meteoroids are incoming.
Meteoric dust produces the silicate features in atmospheric transmission spectra that would be able to diagnose whether the meteoric dust abundance is significant (Section \ref{sec:spectrum}).

\item Dusty outflows efficiently obscure the atmospheric features in transmission spectra, in agreement with the observations of super-puffs (Section \ref{sec:spectrum}).   
Since only tiny dust can be blown up, transmission spectra of dusty outflows tend to exhibit spectral slopes in a wide range of wavelengths.
The radii of super-puffs would decrease with increasing the wavelength if the atmospheric dust causes the radius enhancement, which can be tested by future observations of JWST.
In addition, we find that dusty outflows potentially cause the ``cloud-base effect'' when the dust is completely blown up by an outflow, which might be useful to examine whether or not the dusty outflow is present.

\item We have discussed why super-puffs are uncommon despite the suggested universality of photochemical hazes (Section \ref{sec:MR}).
Using an interior structure model, we have demonstrated that the radius enhancement by atmospheric dust can work only when the planet is near the evaporation limit, regardless of whether or not the dust can be transported to the upper atmosphere.
This is because the drastic radius enhancement is viable only when the pressure scale height is so large that the restricted Jeans parameter becomes $\Lambda\la20$ (Equation \ref{eq:main_ratio}).
Such a large scale height always causes intense atmospheric escape by increasing the sonic point density.
We have predicted that only planets on the verge of total atmospheric loss, corresponding to the mass range of $M_{\rm p}{\sim}2$--$5M_{\rm \oplus}(T_{\rm eq}/500~{\rm K})^{4/3}$ (Equation \ref{eq:mass_range}), can be super-puffs through the radius enhancement caused by high-altitude hazes.

\item Porosity evolution of atmospheric dust hardly affects the qualitative conclusion of this study (Section \ref{sec:porosity}).
Particle growth and subsequent settling still regulate how much dust can be transported by the outflow.
This is because the growth timescale via settling-driven collisions is independent of particle density (Equation \ref{eq:tau_coal}).
However, we found that atmospheric dust potentially grows into aggregates with relatively compressed internal structures in escaping atmospheres, which may act to flatten the observed transmission spectra.

\item Transmission spectra would be useful to study whether the large radius of a super-puff is caused by atmospheric dust or circumplanetary ring (Section \ref{sec:ring}).
In contrast to the atmospheric dust scenario, the observable radius is less dependent on wavelength because circumplanetary rings are likely composed of huge particles ($\ga100~{\rm cm}$).
Future observations would be able to distinguish the atmospheric dust scenario and the ring scenario by examining whether or not the observed radius monotonically decreases with increasing wavelength.

\end{enumerate}

Observations of possible escaping atmospheres of super-puffs would provide additional constraints on the dusty outflow.
The degree of atmospheric escape is potentially accessible by observations of the hydrogen Lyman-$\alpha$ line \citep[e.g.,][]{Vidal-Madjar+03,Lecavelier+10,Ehrenreich+15,Bourrier+16} and metastable He $10830\AA$ line  \citep[e.g.,][]{Oklopcic&Hirata18,Spake+18,Allart+18,Mansfield+18,Bourrier+18,Wang&Dai20,Wang&Dai20b}.
Since the radius enhancement depends on both dust and outflow properties, constraints on the degree of atmospheric escape would help to figure out how atmospheric dust forms and evolves on super-puffs.

The presence of dust in the outflow likely affects radiative transfer in the atmosphere, which we have not included in the present study.
For example, the dust acts to heat the upper atmospheres while it cools the lower atmospheres, though this trend depends on dust optical properties \citep[][]{Heng+12,Morley+15,Poser+19,Molaverdikhani+20,Lavvas&Arfaux21}.
Heating sources in the upper atmosphere play an important role in maintaining the outflow \citep{Wang&Dai18}.
Alternatively, while we have suggested that the photochemical haze is a promising candidate as a source of a dusty outflow, the outflow may hinder the production of hazes by shielding the atmosphere from stellar UV photons.
In that case, the dusty outflow may exhibit a time variability: the haze production ceases when the particles are blown up, and then the production is revived once the hazes are depleted in the outflow.
Since the time variability of a dusty outflow potentially affects observable radii, it might be interesting to see the repeatability of the observed radii of super-puffs.

Our study may also provide some insights on the origin of the dust tails around ultra-short period disintegrating planets.
A few Kepler planets exhibit asymmetric, orbit-varying transit light curves that indicate the presence of comet-like tails \citep[][]{Rappaport+12,Rappaport+14,Sanchis-Ojeda+15}.
These planets undergo intense stellar irradiation that vaporizes rocky surfaces and drives the escape of mineral atmospheres entraining recondensed dust \citep{Perez-Becker&Chiang13,Kang+21}.
The dust tail offers clues to the rocky composition of planets through spectroscopy with future space-based telescopes \citep{Bodman+18,Okuya+20}, whereas the origin of the tail is still highly uncertain. 
\citet{Rappaport+12} estimated the dust-to-gas mass ratio of order unity in the outflow.
If this is true, a simple extrapolation of our results suggests that the recondensed dust should be formed near or beyond the sonic point; otherwise, dust quickly coagulates with each other and settles down to the planet.
In reality, the outflows on disintegrating planets are highly asymmetric \citep{Kang+21}, and further modeling efforts will be warranted to better understand the origin of the dust tails.

%%%%%%%%%%%%%%%%%%%%%%%%%%%%%%%%%%
\acknowledgments 
We thank Xi Zhang and Peter Gao for sharing the draft of \citet{Gao&Zhang20} with us and fruitful conversations.
We also thank an anonymous reviewer for a number of constructive comments, Jonathan Fortney for giving feedback on our early draft, Kenji Kurosaki and Tristan Guillot for enlightening discussions on interior structure modeling, Panayotis Lavvas for sharing the soot refractive indices with us, and Yui Kawashima, Yuichi Ito, Hiroyuki Kurokawa, Yasunori Hori, and Riouhei Nakatani for helpful comments.
This work is supported by JSPS KAKENHI Grant Nos. JP18J14557, JP18H05438, and JP19K03926. 
K.O. acknowledges support from the JSPS Overseas Research Fellowships and use of the lux supercomputer at UC Santa Cruz, funded by NSF MRI grant AST 1828315.

\appendix
%%%%%%%%%%%%%%%%%%%%%%%%%%%%
\section{Analytic Estimation of the Dust Abundance in Escaping Atmospheres}\label{sec:anal_derivation0}
In this Appendix, we construct a simple analytical theory that predicts the dust abundance in escaping atmospheres in an order-of-magnitude sense.
To this end, we evaluate how the upward dust mass flux evolves in the outflow.
The dust production may be negligible above the production height.
In that case, Equation \eqref{eq:basic} indicates that a steady-state size distribution follows
\begin{equation}\label{eq:appendix1}
    %\frac{1}{2}\int_{\rm 0}^{m}K(m',m-m')n(m')n(m-m')dm'- n(m)\int_{\rm 0}^{\infty}K(m,m')n(m')dm'+\frac{1}{r^2}\frac{\partial}{\partial r}\left[ r^2\left(\rho_{\rm g}K_{\rm z}\frac{\partial}{\partial r}\left( \frac{n(m)}{\rho_{\rm g}}\right)-(v_{\rm g}-v_{\rm d})n(m)\right)\right]+S(m,P)=0
    %-\frac{1}{2}\int_{\rm 0}^{m}K(m',m-m')n(m')n(m-m')dm'+ n(m)\int_{\rm 0}^{\infty}K(m,m')n(m')dm'+\frac{1}{r^2}\frac{\partial}{\partial r}\left[ r^2\left((v_{\rm g}-v_{\rm d})n(m)\right)\right]=-\frac{1}{4\pi r^2m}\frac{\partial}{\partial m}\left(\frac{\partial \dot{M}_{\rm dust}}{\partial r}\right),
    \frac{1}{r^2}\frac{\partial}{\partial r}\left[ r^2\left(~(v_{\rm g}-v_{\rm d})~n(m)~\right)\right]=    \frac{1}{2}\int_{\rm 0}^{m}K(m',m-m')~n(m')~n(m-m')~dm'- n(m)\int_{\rm 0}^{\infty}K(m,m')~n(m')~dm'
\end{equation}
where we omit the eddy diffusion for the sake of simplicity.
Multiplying $m$ and integrating from $m=0$ to $m_{\rm cri}$, where $m_{\rm cri}=4\pi a_{\rm cri}^3\rho_{\rm p}/3$, the evolution equation of upward dust mass flux $F$ is given by
\begin{equation}\label{eq:appendix_transport}
     \frac{1}{r^2}\frac{\partial F}{\partial r}=\frac{1}{2}\int_{\rm 0}^{m_{\rm cri}}m~dm\int_{\rm 0}^{m}K(m',m-m')~n(m')~n(m-m')~dm'- \int_{\rm 0}^{m_{\rm cri}}m~n(m)~dm\int_{\rm 0}^{\infty}K(m,m')~n(m')~dm',
\end{equation}
This equation means that the upward mass flux decreases with increasing altitude through collision growth followed by gravitational settling.
The particle mass is largely concentrated at a characteristic mass of $m_{\rm *}\sim m_{\rm cri}$ (Figure \ref{fig:dust_distribution}).
Hence, the size distribution may be crudely approximated by the delta function $n(m)\approx N_{\rm *}\delta(m-m_{\rm cri})$, where $N_{\rm *}$ is the number density.
Then, Equation \eqref{eq:appendix_transport} reduces to
\begin{equation}
    %\frac{1}{r^2}\frac{\partial F}{\partial r}=-\frac{1}{2}m_{\rm *}K_{\rm *}N_{\rm *}^2
    \frac{\partial F}{\partial r}\approx -\frac{1}{2}\frac{K_{\rm *}}{r^2m_{\rm cri}v_{\rm g}^2}F^2,
\end{equation}
where $K_{\rm *}=K(m_{\rm cri},m_{\rm cri})$, and we have used a relation of $F\approx r^2m_{\rm *}N_{\rm *}v_{\rm g}$.
The solution of this equation may be approximated by
\begin{equation}
    \frac{1}{F}\approx \frac{1}{F_{\rm 0}}+\frac{K_{\rm *}\delta r}{2r^2m_{\rm cri}v_{\rm g}^2},
\end{equation}
where $\delta r$ is the length scale of the flux decaying, which we assume is the pressure scale height (i.e., $\delta r=H$).
$F_{\rm 0}$ is the integration constant.
Since the upward flux approaches $F=f_{\rm dust}r^2 \rho_{\rm g}v_{\rm g}$ when the growth is negligible, $F_{\rm 0}=f_{\rm dust}r^2 \rho_{\rm g}v_{\rm g}$.
Therefore, the upward dust mass flux is evaluated as
\begin{equation}\label{eq:F_appendix}
    F\approx \frac{f_{\rm dust}r^2\rho_{\rm g}v_{\rm g}}{1+\chi},
\end{equation}
where we have introduced a dimensionless parameter $\chi$ defined as
\begin{equation}\label{eq:chi_appro0}
    \chi = \frac{K_{\rm *}\rho_{\rm g}H}{2m_{\rm cri}v_{\rm g}}f_{\rm dust}.
\end{equation}
The physical meaning of $\chi$ is the ratio of the transport timescale by atmospheric outflow $H/v_{\rm g}$ to the collision growth timescale for the dust mass density of $\rho_{\rm g}f_{\rm dust}$.
The upward transport is faster than the growth for $\chi<1$ and vice versa for $\chi>1$.
Equation \eqref{eq:F_appendix} can also be written in terms of the dust mass mixing ratio as
\begin{equation}\label{eq:wd_appendix}
    w_{\rm d}\approx \frac{f_{\rm dust}}{1+\chi}.
\end{equation}
%It is obvious that the dust abundance approaches to $f_{\rm dust}$ in the limit of $\chi\ll1$.

In practice, one needs to evaluate $\chi$ at a certain altitude.
Since the particle growth mainly occurs at the dust production region, we evaluate $\chi$ at $P=P_{\rm 0}$.
For similar-sized collisions, we approximate the collision Kernel as
\begin{eqnarray}\label{eq:K_single}
    \nonumber
    K_{\rm *}&=&4\pi a_{\rm cri}^2\left( \epsilon v_{\rm t}(a_{\rm cri})^2+\frac{16k_{\rm B}T}{\pi m_{\rm cri}}\right)^{1/2}\\
    &=&\frac{3}{2}\sqrt{\frac{\pi}{2}}\epsilon\frac{GM_{\rm p}m_{\rm cri}}{\rho_{\rm g}r^2c_{\rm s}}\left( 1+\frac{16k_{\rm B}T}{\pi \epsilon^2 m_{\rm cri}v_{\rm g}^2}\right)^{1/2},
\end{eqnarray}
where we use $v_{\rm t}(a_{\rm cri})=v_{\rm g}$ in the bracket.
We have introduced an order-of-unity constant of $\epsilon$ to account for the width of the size distribution that causes a nonzero relative velocity for the settling-driven collisions \citep[e.g.,][]{Ohno&Okuzumi18,Ormel&Min19}.
We found that $\epsilon=0.25$ leads the theory to well predict the dust abundance.
Inserting Equation \eqref{eq:K_single} into \eqref{eq:chi_appro0}, the parameter $\chi$ can be written as
\begin{equation}\label{eq:chi_appro}
    \chi = \frac{3}{16}\sqrt{\frac{\pi}{2}}\frac{f_{\rm dust}}{\mathcal{M}}\left( 1+\frac{256m_{\rm g}}{\pi m_{\rm cri}}\mathcal{M}^{-2}\right)^{1/2},
    %= \frac{3}{8}\sqrt{\frac{\pi}{2}}\frac{c_{\rm s}}{v_{\rm g}(P_{\rm 0})}f_{\rm dust}\left( 1+\frac{64k_{\rm B}T}{\pi m_{\rm cri}v_{\rm g}(P_{\rm 0})^2}\right)^{1/2}.
\end{equation}
where $v_{\rm g}/c_{\rm s}$ is the Mach number at the dust production height.
One can evaluate the dust abundance in the escaping atmosphere by combining Equations \eqref{eq:wd_appendix} and \eqref{eq:chi_appro}.
It is worth seeing an asymptotic behavior of the dust abundance, given by
\begin{eqnarray}
  w_{\rm d} \approx \left\{ \begin{array}{ll}
    {\displaystyle f_{\rm dust}} & (\chi \ll 1) \\
   {\displaystyle 4.25 \mathcal{M} \left( 1+\frac{256m_{\rm g}}{\pi m_{\rm cri}}\mathcal{M}^{-2}\right)^{-1/2}} & (\chi \gg 1)
  \end{array} \right.
\end{eqnarray}
In the limit of $\chi\ll1$, where the growth is negligible, the dust abundance approaches to $f_{\rm dust}$. 
On the other hand, in the opposite limit of $\chi\gg1$, the abundance approximately plateaus at $w_{\rm d}\sim \mathcal{M}$ when the Mach number is relatively high.
The abundance is further reduced when the Mach number is very small, relevant to the lower atmosphere or a long mass-loss timescale.
Our derivation made several simplifications, such as a monodisperse size distribution.
Nevertheless, as shown in the main text, our theory (Equations \eqref{eq:wd_appendix} and \eqref{eq:chi_appro}) does a reasonable job of predicting the dust abundance in escaping atmospheres.

%%%%%%%%%%%%%%%%%%%%%%%%%%%%%%%%%%%%%%%%%
\section{Description of the Interior Structure Model}\label{appendix:MR}
In this Appendix, we describe the interior structure model used in Section \ref{sec:MR}.
The model is based on a set of stellar structure equations, namely mass conservation, hydrostatic balance, and thermodynamic equations:
\begin{equation}\label{eq:interior1}
    %\frac{\partial r}{\partial M_{\rm r}}=\frac{1}{4\pi r^2 \rho},
    \frac{\partial M}{\partial r}=4\pi r^2 \rho,
\end{equation}
\begin{equation}\label{eq:interior2}
    \frac{\partial P}{\partial r}=-\frac{GM\rho}{r^2},
\end{equation}
\begin{equation}\label{eq:interior3}
    \frac{\partial T}{\partial r}=-\frac{GM\rho}{r^2}\frac{T}{P}\nabla,
\end{equation}
where $M$ is the mass enclosed in a sphere with a radius of $r$.
Assuming a homogeneously mixed atmosphere for the sake of simplicity, the temperature gradient $\nabla\equiv d\log{T}/d\log{P}$ is determined from the Schwarzschild criterion, which reads
\begin{equation}
    \nabla = \min{(\nabla_{\rm ad},\nabla_{\rm rad})},
\end{equation}
where $\nabla_{\rm ad}$ and $\nabla_{\rm rad}$ are the adiabatic and radiative temperature gradient.
In this study, we simply assume the ideal dry gas for which the adiabatic gradient is $\nabla_{\rm ad}=(\gamma-1)/\gamma$, where $\gamma$ is the adiabatic index and set to $\gamma=7/5$ for diatomic gasses.
Strictly speaking, one needs a radiative transfer model to evaluate the radiative gradient \citep[e.g.,][]{Fortney+07}.
Instead, we evaluate the radiative gradient by differentiating the analytical solution of \citet{Guillot10} for a radiative atmosphere with a double-gray approximation (their Equation (29)), given by
\begin{equation}\label{eq:nabla_rad}
    \nabla_{\rm rad}=-\frac{3\kappa L_{\rm int}}{64\pi \sigma_{\rm SB}GM}\frac{P}{T^4}\left[ 1+4f\left( \frac{T_{\rm eq}}{T_{\rm int}}\right)^4(1-\gamma_{\rm opa}^2)\exp{(-\gamma_{\rm opa}\sqrt{3}\tau)}\right],
\end{equation}
where $L_{\rm int}=4\pi r^2 \sigma_{\rm SB}T_{\rm int}^4$ is the planetary intrinsic luminosity evaluated at the radiative-convective boundary, $f=1/4$ is the redistribution factor, which we assume for an averaging over the whole planetary surface, the $\gamma_{\rm opa}$ is the visible to thermal opacity ratio, $\kappa$ is the thermal opacity which is taken from the analytical fit of Rosseland mean opacity provided by \citet{Freedman+14}. 
We assume $\gamma_{\rm opa}=0.032$ that reasonably reproduces the temperature structure of GJ1214b, whose equilibrium temperature ($T_{\rm eq}=530~{\rm K}$) is comparable to those of typical super-puffs, computed by a radiative transfer model \citep{Valencia+13}.
The intrinsic luminosity depends on the thermal evolution history, but we vary $T_{\rm int}$ as a free parameter for sensitivity check in this study.
Note that Equation \eqref{eq:nabla_rad} reduces to the radiative diffusion approximation used for self-luminous objects in the limit of either $\tau \gg 1/\sqrt{3}\gamma_{\rm opa}$ or $T_{\rm int}\gg T_{\rm eq}$.
$\tau$ is the thermal optical depth calculated by
\begin{equation}\label{eq:interior4}
    \frac{\partial \tau}{\partial r}=\rho \kappa.
\end{equation}
Equations \eqref{eq:interior1}, \eqref{eq:interior2}, \eqref{eq:interior3}, and \eqref{eq:interior4} involve five variables ($M$, $P$, $\rho$, $T$, $\tau$), and thus we need an additional constrain to close the equations, which is provided by the equation of state (EOS).
In this study, we simply adopt the ideal gas EOS of $P=\rho k_{\rm B}T/m_{\rm g}$ for solar composition atmosphere with $m_{\rm g}=2.35~{\rm amu}$.
We do not compute the interior structure of the core and instead set the inner boundary condition at core surface, which is given by \citep{Zeng+19}
\begin{equation}
        R_{\rm core}=R_{\rm \oplus}\left( \frac{M_{\rm core}}{M_{\rm \oplus}}\right)^{1/3.7}(1+0.55x_{\rm ice}-0.14x_{\rm ice}^2),
\end{equation}
where $x_{\rm ice}$ is the ice mass fraction of the core which we set to zero.
The boundary condition is given by $M(r_{\rm s})=M_{\rm p}$, $M(R_{\rm core})=M_{\rm core}$, $T(r_{\rm s})=T_{\rm eq}$, $\tau(r_{\rm s})=0$.
An outer boundary condition is in prior unknown for $P$.
For given atmospheric mass fraction $f_{\rm atm}=1-M_{\rm corre}/M_{\rm p}$ and intrinsic temperature $T_{\rm int}$, we find out the proper boundary condition of $P$ by using the shooting method.

\end{document}